\documentclass[review,authoryear,12pt]{elsarticle}
\usepackage{amsmath,bm}
\usepackage{makecell}
\usepackage{algorithm}
\usepackage{algpseudocode}
\usepackage{amssymb}
\usepackage{amsmath}
\usepackage{placeins}
\usepackage{latexsym}
\usepackage{amsmath}
\usepackage{amssymb}
\usepackage{subfig}
\usepackage{float}
\usepackage{tikz,tkz-tab}
\usepackage{textcomp}
\graphicspath{{pictures/}}
\usepackage{longtable}
\usepackage{multirow}
\usepackage[utf8]{inputenc}
\usepackage{lscape}
\usepackage[hidelinks]{hyperref}
\usepackage{caption}
\usepackage{mathrsfs}

\begin{document}

\begin{frontmatter}

\title{Generative Parametric Design (GPD): A framework for real-time geometry generation and on-the-fly multiparametric approximation.}

\author[label1]{M. El Fallaki Idrissi\corref{cor1}}
\cortext[cor1]{Corresponding author.}\ead{mohammed.el\_fallaki\_idrissi@ensam.eu}

\author[label1]{J. Mounayer}
\author[label1]{S. Rodriguez}
\author[label2]{F. Meraghni}
\author[label1]{F. Chinesta}

\address[label1]{Arts et Métiers Institute of Technology, CNRS, PIMM-UMR 8006, 151 Boulevard de l’Hôpital, 75013 Paris, France.}
\address[label2]{Arts et Métiers Institute of Technology, CNRS, LEM3-UMR 7239, 4 rue Augustin Fresnel, 57078 Metz, France}

\begin{abstract}

This paper presents a novel paradigm in simulation-based engineering sciences by introducing a new framework called Generative Parametric Design (GPD). The GPD framework enables the generation of new designs along with their corresponding parametric solutions given as a reduced basis. To achieve this, two Rank Reduction Autoencoders (RRAEs) are employed, one for encoding and generating the design or geometry, and the other for encoding the sparse Proper Generalized Decomposition (sPGD) mode solutions. These models are linked in the latent space using regression techniques, allowing efficient transitions between design and their associated sPGD modes. By empowering design exploration and optimization, this framework also advances digital and hybrid twin development, enhancing predictive modeling and real-time decision-making in engineering applications. The developed framework is demonstrated on two-phase microstructures, in which the multiparametric solutions account for variations in two key material parameters.
\end{abstract}


\end{frontmatter}

\FloatBarrier
\section{Introduction}
Today, designers, engineers, and manufacturers must navigate the complexities of developing highly customizable products that not only meet stringent technical constraints but also align with evolving aesthetic trends and market demands. In this dynamic landscape, three fundamental principles have become essential in engineering science: accuracy, speed, and generativity, all of which are critical for developing new and optimized solutions that effectively meet evolving requirements.

Over the past decades, achieving high accuracy has been a central goal in engineering sciences and modeling \citep{akbar2016modeling, rappaz2003numerical}. Advances in numerical methods, physics-based simulations, and computer science have enabled reliable predictions of real-world product behavior \citep{schafer2006computational, dhatt2012finite,wriggers2024virtual, chatzigeorgiou2022multiscale}. This ensures that structures consistently meet stringent performance, safety, and durability standards under various operating conditions, including temperature fluctuations, humidity, and loading. However, attaining high fidelity simulations comes at a significant computational cost. This computational cost hinders the iterative design process, making it difficult to incorporate into real-time decision-making or rapid prototyping workflows.

In modern engineering, accuracy alone is no longer sufficient, speed has become equally critical especially in an era of rapid technological advancement and on-demand manufacturing. Fast simulations and real-time responses are essential for real-time interactions between users and models, such as rapid prototyping, digital twins, and interactive simulations. To meet these demands, Model Order Reduction (MOR) techniques and machine learning based surrogate models have emerged as powerful solutions for accelerating simulations and reducing the complexity of traditional modeling approaches \citep{lassila2014model, chinesta2011short, schilders2008introduction, chen2025physics, wu2023deep}. This is especially important when dealing with high-dimensional parametric problems. In this context, the Proper Generalized Decomposition (PGD) has proven highly effective in addressing numerous challenges in engineering sciences, particularly in solving high-dimensional and multiparametric problems \citep{chinesta2010recent,idrissi2025advanced, idrissi2023non, idrissi2022multiparametric, sancarlos2021pgd, pasquale2022parametric}. It excels at solving high-dimensional and multiparametric problems, including scalar, curve, and field-based problems, by employing a separated representation of the solution, where variables such as space, time, loading conditions, and material properties are decoupled. Instead of solving the problem in a high-dimensional space, PGD reformulates it into a series of lower-dimensional problems, significantly reducing computational complexity. This approach enables the generation of computational vademecums and virtual charts, which facilitate real-time simulation and optimization of complex problems \citep{chinesta2013pgd, ghnatios2012proper}. Such capabilities are essential in advancing digital, virtual, and hybrid twin paradigms \citep{chinesta2020virtual, rodriguez2023hybrid}. 

Beyond accuracy and speed, the concept of generativity is transforming the design process. Rather than merely refining existing solutions, generative design methodologies leverage advanced computational techniques to explore vast design spaces, creating entirely new, optimized solutions. By employing machine learning, artificial intelligence, and generative design algorithms, engineers can automatically generate and evaluate thousands of potential configurations, balancing performance, efficiency, and manufacturability, even in non-parametric settings. In recent years, various generative approaches have been developed, with Autoencoders (AEs) \citep{bank2023autoencoders}, Variational Autoencoders (VAEs) \citep{doersch2016tutorial} and Generative Adversarial Networks (GANs) \citep{creswell2018generative} being among the most popular models 
\citep{Regenwetter2022}. However, the adoption of generative models in engineering design applications remains relatively limited, primarily due to the need for large, high-quality training datasets. Such datasets are often proprietary, confidential, and difficult to access in industrial contexts, which constrains their availability for research and practical applications. Furthermore, in engineering applications, generating a design alone is insufficient; the proposed solution must also be analyzed and validated to ensure its feasibility and performance. This requires integrating generative models with high-fidelity simulation tools, optimization algorithms, and physical constraints to bridge the gap between conceptual design and functional implementation. Consequently, only a few studies have addressed these challenges \citep{Oh2019,Yoo2021}.  

In real-world applications, designs and geometries are often represented in complex, high-dimensional spaces, making their control and manipulation challenging. A crucial step in improving generative design models is developing methods to map the high-dimensional design space into a well-structured and regularized latent space that accurately represents the geometry while remaining intuitive and interpretable for engineers and designers. Recently, a novel method called Rank Reduction Autoencoders (RRAEs) has been developed \citep{mounayer2024rank}, which combines the regularization of truncated Singular Value Decomposition (SVD) with the nonlinear feature extraction capabilities of autoencoders. This approach has demonstrated promising results in addressing challenges frequently encountered in engineering applications, particularly in the interpolation of complex physical phenomena \citep{elfallaki2025a, tierz2025v}. Compared to traditional techniques like standard AEs and VAEs, RRAEs and their variants such as variational RRAEs (VRRAEs) \citep{mounayer2025variational} and adaptive RRAEs (aRRAEs) \citep{mounayer2024rank}, demonstrate more robust performance, especially in constructing compact and well-regularized latent spaces of high-dimensional data.

In the face of increasingly complex engineering challenges, the need to balance simulation accuracy, computational efficiency, and generative design capabilities has become more critical than ever. This challenge is particularly pronounced in scenarios where data is scarce, and problems exist within high-dimensional spaces. To address these challenges, we introduce a novel framework called Generative Parametric Design (GPD), which enables the real-time generation of new designs while simultaneously constructing their corresponding multiparametric solutions in the form of a MOR. This is achieved by employing two RRAEs models: one for encoding and generating the geometry, and the other for encoding the PGD modes decomposition of the multiparametric solutions. By linking these models in the latent space through regression techniques, the framework enables the prediction of PGD modes for each generated geometry or design. This framework is a useful tool for design exploration and optimization, allowing the generation of new geometries and the adjustment of parameters like material properties, boundary conditions, and loads, while predicting system responses in real-time.


The structure of this paper is as follows: Section \ref{sec2:Generative Parametric Design} introduces the Generative Parametric Design framework, outlining the core methodologies of Rank Reduction Autoencoders and Proper Generalized Decomposition that form its foundation. Section \ref{sec3: application} presents the application of the GPD framework to microstructures, demonstrating its effectiveness in designing new microstructural configurations and constructing their associated multi-parametric solutions for microscale fields governed by material parameters. This section also includes a detailed analysis of the results, assessing the accuracy and performance of the proposed method in comparison to reference solutions. Finally, Section \ref{sec4: Conclusion} summarizes the main findings and discusses potential directions for future research.

\FloatBarrier
\section{Generative Parametric Design: A new Framework }
\label{sec2:Generative Parametric Design}
\begin{figure}[tbh!]
	\centering
	\includegraphics[width=0.7\textwidth]{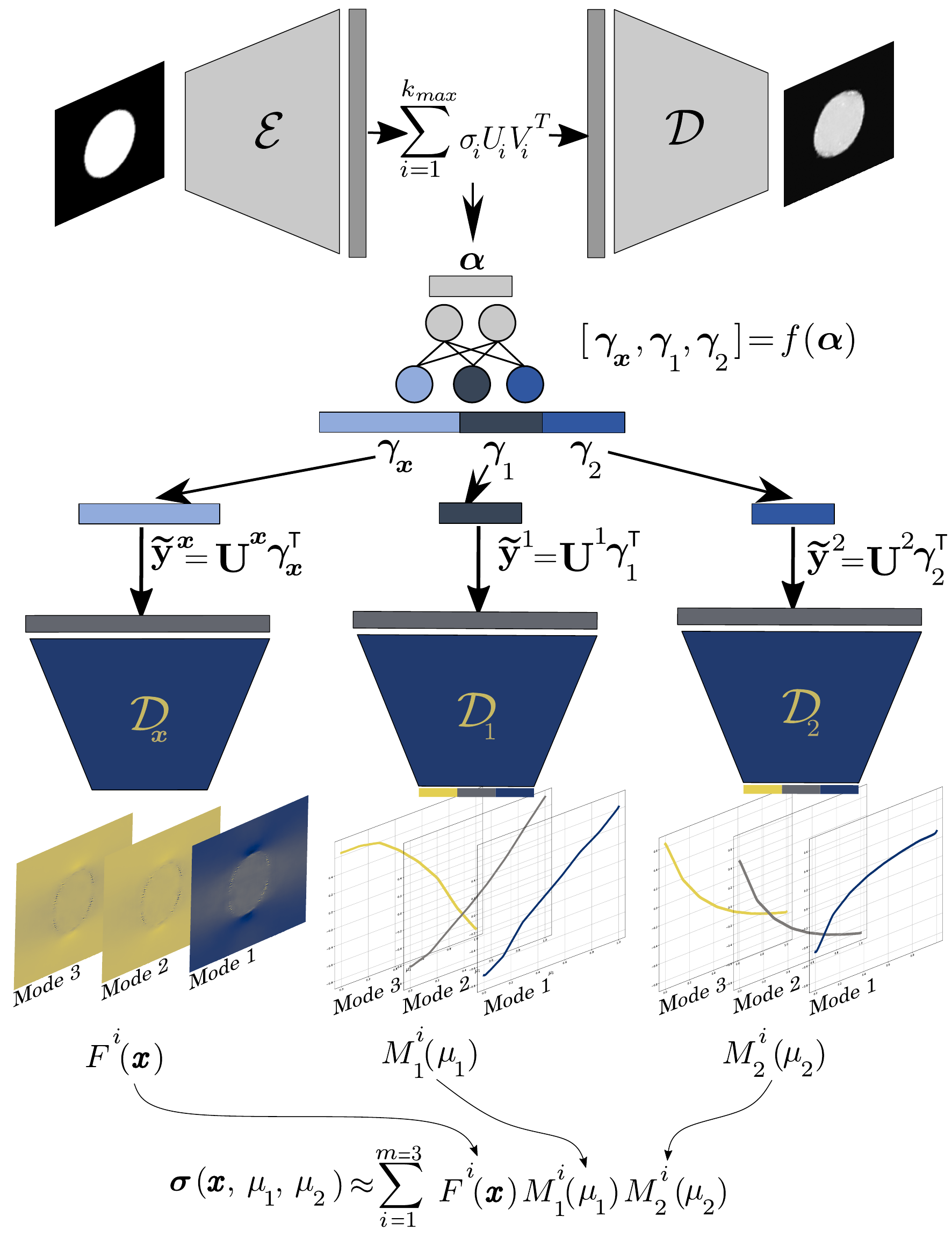}
	\caption{Schematic illustration of the Generative Parametric Design (GPD) concept. After separately training the four RRAE models and computing the corresponding latent coefficients for each sample, a multi-layer perceptron (MLP) is trained using this dataset of input-output pairs (Section \ref{sec3: application}). Once trained, the framework supports two primary workflows: (i) given a microstructure morphology, the geometry encoder $(\mathcal{E})$ maps it to a latent vector $(\boldsymbol{\alpha})$, the MLP predicts the latent coefficients of the PGD solution modes $([\boldsymbol{\gamma}_{\boldsymbol{x}}, \boldsymbol{\gamma}_{1}, \boldsymbol{\gamma}_{2}])$, and the decoders $(\mathcal{D}_{\boldsymbol{x}}, \mathcal{D}_{1}, \mathcal{D}_{2})$ reconstruct the full multiparametric solution; (ii) new designs can be generated by sampling the geometry latent space $(\boldsymbol{\alpha})$, which is decoded by $(\mathcal{D})$ produce the generated microstructure, and passed through the MLP–decoder pipeline to reconstruct the corresponding PGD solution.}
	\label{fig:Generative_Parametric_Design}
\end{figure}

In this work, we introduce a novel framework called \textbf{Generative Parametric Design} as shown in Figure \ref{fig:Generative_Parametric_Design}. The term \textit{Generative} refers to the ability to create new designs and geometries, by interpolating or sampling within a regularized latent space learned by RRAEs model. Meanwhile, the term \textit{Parametric} highlights the capability to predict multi-parametric solutions, in real-time, particularly the PGD modes, by incorporating parameters such as material properties, loading, and boundary conditions. This means that for each newly generated design, a corresponding multi-parametric solution approximation is dynamically constructed \textit{on-the-fly}. The GPD framework combines the strengths of RRAEs and the PGD methods to efficiently approximate complex, multi-parametric solutions while exploring new design. RRAEs are used to map high-dimensional data, such as curves and images, into a structured, low-dimensional, and regularized latent space. At the same time, PGD provides a reduced representation of the solution in terms of separated modes, allowing the construction of multiparametric solutions with limited datasets. Within this framework, both the geometry/design and the PGD solution modes are mapped into their respective latent spaces, $\{\boldsymbol{\alpha}\}$, and $\{\boldsymbol{\gamma}_{\boldsymbol{x}}, \boldsymbol{\gamma}_{1}, \boldsymbol{\gamma}_{2} \}$. A regression model is then trained to map the geometry latent representation to that of the PGD modes, enabling direct prediction of solution modes from a given geometry. This capability is particularly valuable in industrial settings, where geometries are often complex, parametric spaces are large and only sparse solution data are available. The underlying RRAEs and PGD methodologies are briefly reviewed in the following subsections.

Figure \ref{fig:Generative_Parametric_Design} illustrates the GPD framework for online use once it has been trained. The framework operates along two distinct workflows. In the first workflow, a given microstructure morphology is encoded by the geometry encoder $\mathcal{E}$ to obtain latent coefficients $\boldsymbol{\alpha}$. An MLP then maps $\boldsymbol{\alpha}$ to the latent coefficients of the PGD solution modes, $[\boldsymbol{\gamma}_{x}, \boldsymbol{\gamma}_{1}, \boldsymbol{\gamma}_{2}]$. The trained decoders $\mathcal{D}_{x}$, $\mathcal{D}_{1}$ and $\mathcal{D}_{2}$ subsequently reconstruct the corresponding PGD modes, which are combined to produce the full multiparametric solution. This enables real-time approximation of multi-parametric responses for a given microstructure. In the second workflow, new microstructures are generated by sampling within the geometry latent space  $\boldsymbol{\alpha}$. These sampled latent vectors are optionally decoded by the geometry decoder $\mathcal{D}$ to produce novel microstructure images, and the same MLP + PGD-decoder pipeline is used to reconstruct the corresponding multiparametric solution for each new sampled $\boldsymbol{\alpha}$.

\FloatBarrier
\subsection{Rank Reduction AutoEncoders (RRAEs)}
\label{subsec:RRAEs}

\begin{figure}[hbt!]
	\centering	\includegraphics[width=0.8\textwidth]{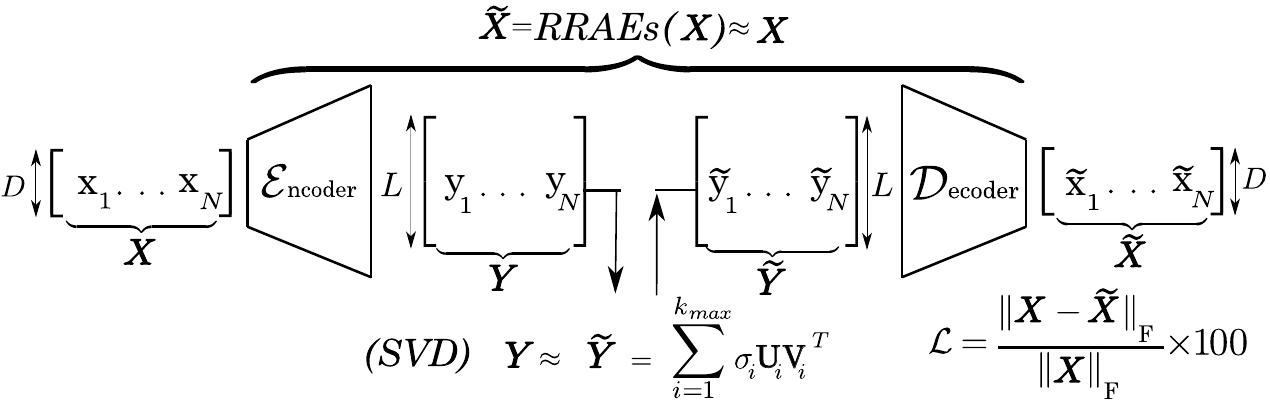}
	\caption{Schematic illustration of the Rank Reduction AutoEncoders (RRAEs) model.}
	\label{fig:RRAEs:model}
\end{figure}

Combining the efficient linear compression capabilities of truncated Singular Value Decomposition (SVD) with the powerful nonlinear feature extraction of autoencoders, a nonlinear dimensionality reduction method, called Rank Reduction Autoencoders (RRAEs), was introduced in \citep{mounayer2024rank}. This method is specifically designed to perform nonlinear dimensionality reduction while enforcing a regularized latent space that preserves the most relevant features needed to accurately represent the original data.


As illustrated in Figure~\ref{fig:RRAEs:model}, the RRAEs model does not compress the input data directly into a compact latent space. Instead, it first maps the input matrix $\boldsymbol{X}= \{\mathbf{x}_1, \ldots, \mathbf{x}_N\} \in \mathbb R^{D\times N}$, where $D$ is the input dimensionality and $N$ is the number of observations, to a high-dimensional latent representation $\boldsymbol{Y} = \{\mathbf{y}_1, \ldots, \mathbf{y}_N\} \in \mathbb R^{L\times N}$ using an encoder function $\boldsymbol{Y} = \mathcal{E}(\boldsymbol{X})$. Here, $L$ is the dimension of the latent space, which is chosen to be larger than the intrinsic parametric space. While $L$ can be smaller or larger than the original input dimension $D$, it must be sufficiently large to capture and flatten the true underlying structure of the data. A low-rank approximation of $\boldsymbol{Y}$ is then computed using truncated SVD keeping only the top $k_{\max}$ components: 
\begin{equation}
	\boldsymbol{Y} \approx \tilde{\boldsymbol{Y}} = \boldsymbol{U} \boldsymbol{\Sigma} \boldsymbol{V}^\top = \sum_{i=1}^{k_{\max}}{\sigma_{i}\mathbf{U}_{i}\mathbf{V}_{i}^\top},
\end{equation}
where $\boldsymbol{U}= [\mathbf{U}_1, \ldots, \mathbf{U}_{k_{\max}}] \in \mathbb{R}^{L \times k_{\max}}$, $\boldsymbol{V}^\top= [\mathbf{V}_1, \ldots, \mathbf{V}_{k_{\max}}]^\top  \in \mathbb{R}^{k_{\max} \times N}$, and $\boldsymbol{\Sigma} \in \mathbb{R}^{k_{\max} \times k_{\max}}$ is a diagonal matrix of singular values. The singular values are sorted in descending order: $\boldsymbol{\Sigma} = \mathrm{diag}(\sigma_1,\dots, \sigma_{k_{\max}} )$. The choice of $k_{\max}$ controls the effective rank of the approximation and is treated as a hyperparameter, set prior to training. Alternatively, an adaptive RRAE model (aRRAEs) can also be used to automatically learn $k_{\max}$ during training, as described in \citep{mounayer2024rank}.

The reconstruction of a latent sample $\tilde{\boldsymbol{Y}}$ after applying the truncated SVD might be expressed also as:
\begin{equation}
	\tilde{\boldsymbol{Y}} = \boldsymbol{U} \boldsymbol{A},
\end{equation}
with $\boldsymbol{A}= \boldsymbol{\Sigma} \boldsymbol{V}^\top =\{\boldsymbol{\alpha}_1, \ldots, \boldsymbol{\alpha}_{N}\} \in \mathbb{R}^{k_{\max}\times N}$. Each $\boldsymbol{\alpha}_i$ is a latent coefficient vector that compactly represents the input $\mathbf{x}_i$ in a lower-dimensional, structured space. In the following subsections, we will refer to these coefficient vectors as $\boldsymbol{\alpha}$ when working with geometry, and as $\boldsymbol{\gamma}$ when referring to PGD modes. 

Finally, the input data is reconstructed by applying the decoder to the low-rank latent representation $\tilde{\boldsymbol{X}} = \mathcal D (\tilde{\boldsymbol{Y}})$. This step produces the final output $\tilde{\boldsymbol{X}}$, which approximates the original input $\boldsymbol{X}$. The reconstruction loss, which quantifies the difference between the input and its reconstruction, is defined as:
\begin{equation}
	\mathcal{L} = \dfrac{||\boldsymbol{X}-\tilde{\boldsymbol{X}}||_F}{|| \boldsymbol{X}||_F}\times 100,
	\label{eq:reconstruction-loss}
\end{equation}
where $||\cdot||_F$ denotes the Frobenius norm calculated between the flattened original and reconstructed vectors, normalized by the magnitude of the ground-truth input.

\FloatBarrier
\subsection{Non intrusive Proper Generalized Decomposition (PGD): sparce PGD (sPGD)}
\label{subsec:sPGD}
Proper Generalized Decomposition (PGD) is a model-order reduction technique based on a separated representation of variables \citep{chinesta2010recent,chinesta2011short,chinesta2011overview}. It expresses the high-dimensional solution as a sum of products of one-dimensional functions, as shown in Equation \ref{Eq:PGD_seperated_representation}. This strategy alleviates the computational burden of high-dimensional problems by enabling efficient approximations. 

In this study, we employ a non-intrusive variant of PGD, known as Sparse PGD (sPGD) \citep{ibanez2018multidimensional}, to construct multi-parametric approximations of the von Mises stress field across different microstructure morphologies. This method efficiently captures the variations in the stress response as the material properties of the matrix ($\mu_1$) and inclusion phases ($\mu_2$) vary, as detailed in the following section. It offers an efficient way to build sparse representations in high-dimensional parametric spaces, while overcoming the effects of the so-called ``\textit{curse of dimensionality}".

For the specific case studied, where two material parameters, $\mu_1$ and $\mu_2$, are varied, the stress field is approximated using a separated representation as follows:
\begin{equation}
	\label{Eq:PGD_seperated_representation}
	\sigma(\boldsymbol{x}, \mu_{1},\mu_{2})\approx\sum_{i=1}^{m}{F^{i}(\boldsymbol{x})M_{1}^{i}(\mu_{1})M_{2}^{i}(\mu_{2})}  
\end{equation}
here, $m$ denotes the number of modes. In this formulation, $F^{i}$ represents the spatial mode that depends on the spatial coordinate $\boldsymbol{x}$, while $M_{1}^{i}$ and $M_{2}^{i}$ are one-directional functions that depend on the parameters $\mu_1$ and $\mu_2$, respectively.

To predict the solution for any new microstructural material configuration, characterized by constitutive materials, without resorting to full numerical simulations, the sPGD method provides an approximate solution $\sigma_{\text{app}}$ using a Galerkin projection approach. This approximate solution is obtained by enforcing the following weak form:
\begin{equation}
	\int_{\Omega_{\mu}}{s^{*}(\boldsymbol{x}, \boldsymbol{\mu})\Big(\sigma_{\text{app}}(\boldsymbol{x}, \boldsymbol{\mu})-\sigma(\boldsymbol{x}, \boldsymbol{\mu})
		\Big) \mathrm{d} \boldsymbol{x}\mathrm{d}\boldsymbol{\mu}=0}
\end{equation}
where $\sigma(\boldsymbol{x}, \boldsymbol{\mu})$ represents the unknown target solution to be approximated from sparse data, and $s^{*}(\boldsymbol{x}, \boldsymbol{\mu}) \in \mathscr{C}^{0}(\Omega_{\mu})$ denotes the test function space. In practice, since the solution is only available at a finite set of sampling points, the test function is constructed as a distribution of Dirac delta functions located at $n$ prescribed parameter samples $\boldsymbol{\mu}^{k} = (\mu_{1}^{k}, \mu_{2}^{k})$, $k = 1, \dots, n$:
\begin{equation}
	s^{*}(\boldsymbol{x}, \boldsymbol{\mu})=\sigma^{*}(\boldsymbol{x}, \boldsymbol{\mu})\sum_{k=1}^{n}{\delta(\boldsymbol{\mu}^{k})} 
	\label{Eq14}
\end{equation}
thereby reducing the Galerkin projection to a collocation-type formulation, consistent with the fact that the solution is known only at those discrete parameter locations.

This formulation can be rewritten as:
\begin{equation}
	s^{*}(\boldsymbol{x},\boldsymbol{\mu})=(F^{m*}M_{1}^{m}M_{2}^{m}+ F^{m}M_{1}^{m*}M_{2}^{m} + F^{m}M_{1}^{m}M_{2}^{m*})\sum_{k=1}^{n}{\delta(\boldsymbol{\mu}^{k})} 
	\label{Eq15}
\end{equation}

Therefore, According to the PGD principle, the solution $\sigma_{\text{app}}(\boldsymbol{x}, \boldsymbol{\mu})$ is approximated by a sum of products of one-dimensional functions:

\begin{equation}
	\sigma_{\text{app}}(\boldsymbol{x},\boldsymbol{\mu}) \approx \sigma^{m}(\boldsymbol{x}, \boldsymbol{\mu}) = \sum_{i=1}^{m}{F^{i}(\boldsymbol{x})M_{1}^{i}(\mu_{1})M_{2}^{i}(\mu_{2}) }
\end{equation}
where $\sigma^{m-1}(\boldsymbol{x}, \boldsymbol{\mu})$ is the known approximation from the previous iteration. At the $m$-th mode, the unknown functions $F^{i}(\boldsymbol{x})$, $M_{1}^{i}(\mu_1)$ and $M_{2}^{i}(\mu_2)$ are determined using a finite element projection and a greedy algorithm. These functions are generally expressed in a parametric form:
\begin{equation}
	M_{j}^{m}(\mu_{j})=\sum_{k=1}^{D_s}{\phi_{k}(\mu_{j})\lambda_{k}} = \boldsymbol{\Phi}_{j}^{T}\boldsymbol{\lambda}_{j}, \,\,\,\, i=1,2
\end{equation}
where contains the degrees of freedom and represents the vector of shape functions with $D_s$ components for the first-th and second-th parameter. A complete mathematical formulation of the sPGD method is provided in the appendix of \citep{rodriguez2023hybrid}.

\FloatBarrier
\section{Application: GPD for two-phases microstructures media}
\label{sec3: application}
For the sake of simplicity and without loss of generality, we demonstrate the feasibility of the proposed Generative Parametric Design framework by applying it to the design of two-phase microstructures. Specifically, we consider Representative Volume Elements (RVEs) composed of inclusions embedded within a matrix. The objective is to leverage the GPD framework to generate novel microstructure designs with varying inclusion shapes, sizes, and orientations. This enables the systematic exploration of a broad design space while simultaneously providing real-time predictions of the corresponding multiparametric responses in a reduced-order form for each generated microstructure.

A variety of design representation methods can be employed to define geometries, surfaces, or solids, depending on the nature of the problem. Common approaches include image-based representations (e.g., pixelization in 2D), voxelization for 3D structures, point clouds, meshes, and graph-based models. Each method offers specific advantages in terms of accuracy, computational efficiency, and suitability for particular design or analysis tasks. In this study, we adopt an image-based representation to define the geometry of the RVEs, using a resolution of 148$\times$148 pixels. 

A dataset of 599 distinct microstructures and their corresponding multiparametric solutions is generated to train and validate the GPD framework. These microstructures include a wide variety of inclusion geometries, with cross-sections varying in shape (rectangular, square, circular, elliptical), size, and orientation. Their mechanical behavior is modeled under the assumption of linear elasticity, with both the matrix and the inclusions treated as isotropic materials. Therefore, two key material parameters are considered: the Young’s modulus of the matrix, defined as $\mu_{1}= E_{1} \in [800, 2400]$ MPa, and that of the inclusion, given by $\mu_{2}= E_{2}\in [12000, 68000]$ MPa. The Poisson’s ratio remains constant for both materials, with $\nu_{1}=0.3$ for the matrix and $\nu_{2}=0.2$ for the inclusion.

\begin{figure}[tbh!]
	\centering	\includegraphics[width=0.5\textwidth]{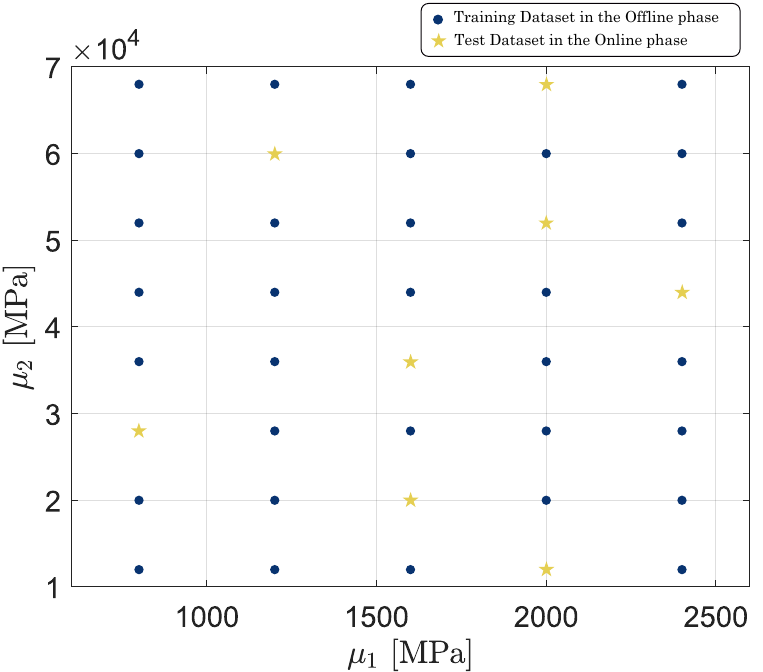}
	\caption{Collocation points employed for constructing and validating the multiparametric solution using the sPGD model.}
	\label{fig:sPGD_grid}
\end{figure}    

For each microstructure, a multiparametric solution is constructed using the sPGD method, as described in the previous section, to predict the local von Mises stress field $\sigma$ under a macroscopic applied strain $\overline{\boldsymbol{\varepsilon}}_{xx}$ of 8$\%$. To achieve this, a regular grid of 40 collocation points is first defined in the 2D material parameter space (corresponding to the Young’s moduli of the matrix and inclusion), as illustrated in Figure \ref{fig:sPGD_grid}. At each collocation point and for each microstructure, the corresponding pair of material parameter pairs is assigned, and a finite-element simulation is performed to extract the von Mises stress field. Of the 40 sample points, 32 are used to build the sPGD approximation and the remaining 8 serve for validation.

For each multiparametric solution, the sPGD is trained using a fixed hyperparameter setting with a $D_s=8$ degrees of freedom and Kriging shape functions for each one-dimensional function. To enforce a uniform data structure for subsequent RRAEs model training, only the first three separated modes were retained for every multiparametric solution, avoiding different number of modes across the different solutions. These three modes were found sufficient to accurately represent the solution, yielding a Mean Absolute Percentage Error (MAPE) below $9 \%$ at the training collocation points and below $8 \%$ at the test points. 

Finally, the complete dataset is split into two subsets. A total of 500 microstructures, along with their sPGD-based multiparametric solutions, are used to train all RRAEs models, learn the latent space representations, and establish the regression mapping between the latent space coefficients. The remaining 99 microstructures are reserved for independent testing and validation, as detailed in the following subsections.

\FloatBarrier
\subsection{RRAEs for learning the morphological features of microstructures}

\begin{figure}[hbt!]
	\centering	\includegraphics[width=\textwidth]{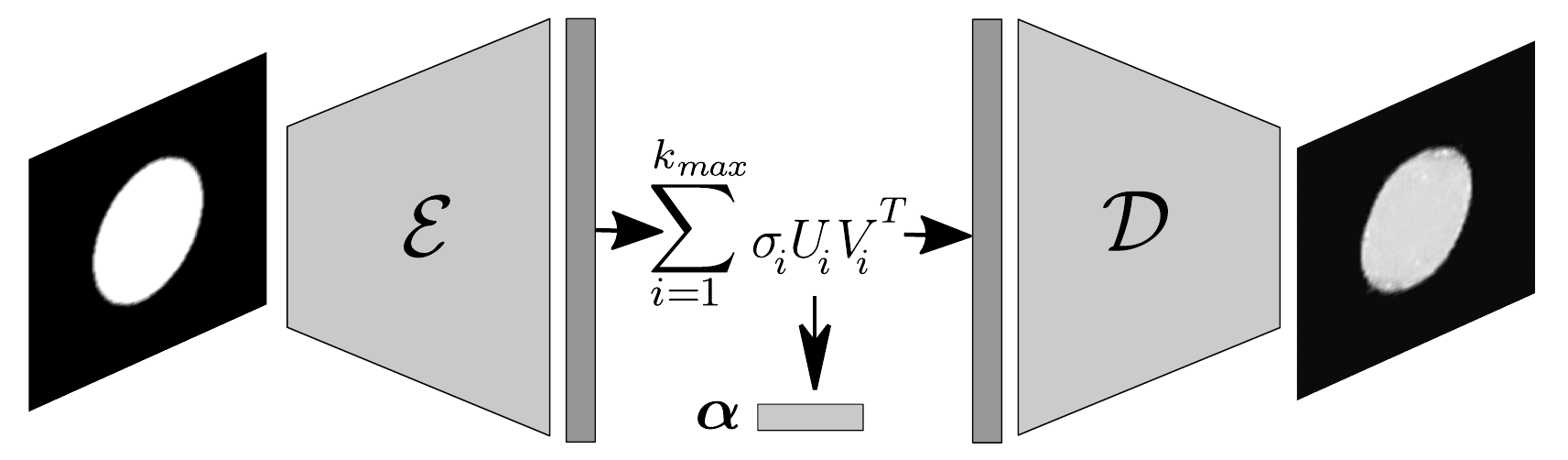}
	\caption{RRAEs model for Geometry (Microstructure Morphology).}
	\label{fig:RRAEs_Geometry:model}
\end{figure}

In this subsection, as illustrated in Figure \ref{fig:RRAEs_Geometry:model}, we present how the RRAEs model captures and represents microstructural morphology. It compresses image-based structural data into a regularized, low-dimensional latent space (Latent space coefficients $\boldsymbol{\alpha}$) that preserves key geometric features with minimal information loss. This compact representation enables accurate reconstruction of the original microstructure, while also supporting smooth interpolation between different designs and the generation of new geometries through latent space sampling.

In this manner and as discussed previously, the dataset is divided into 500 RVEs for training and 99 for testing the developed RRAE model. Approximately 50$\%$ of the samples in each set contain elliptical or circular fibers, while the remaining 50$\%$ include square or rectangular fibers, ensuring a balanced representation of geometric diversity. As mentioned earlier, an image-based representation is adopted to train the RRAEs model for geometries, using a convolutional RRAEs architecture. The input data are formatted as binary images of size  $148\times148$, where the two dimensions correspond to the spatial resolution along the $x$ and $y$-directions. In this representation, pixel values are assigned as 1 for inclusions and 0 for the matrix. Each image is embedded into a latent space of dimension $L=300$, and in that space apply an SVD truncating to $k_{\max}=4$ singular values. A very low $k_{\max}$ can result in poor reconstruction quality, while a higher $k_{\max}$ improves reconstruction accuracy but may reduce the smoothness of interpolation. Following the same hyperparameter settings as in \citep{elfallaki2025a}, the model is trained with a batch size of 20 and a staged learning rate schedule consisting of three phases of 3000 steps each, with learning rates of $10^{-3}$, $10^{-4}$, and $10^{-5}$, respectively. Additional key hyperparameters are provided in Table \ref{tab:RRAEs_hyperparameters}, and further implementation details can be found in \citep{elfallaki2025a}.

\begin{figure}[bht!]
	\centering              
	
	\subfloat[Example from \textbf{training} dataset.]{\label{RRAEs_Geometry:TvsR:Train2}\includegraphics[width=0.45\textwidth]{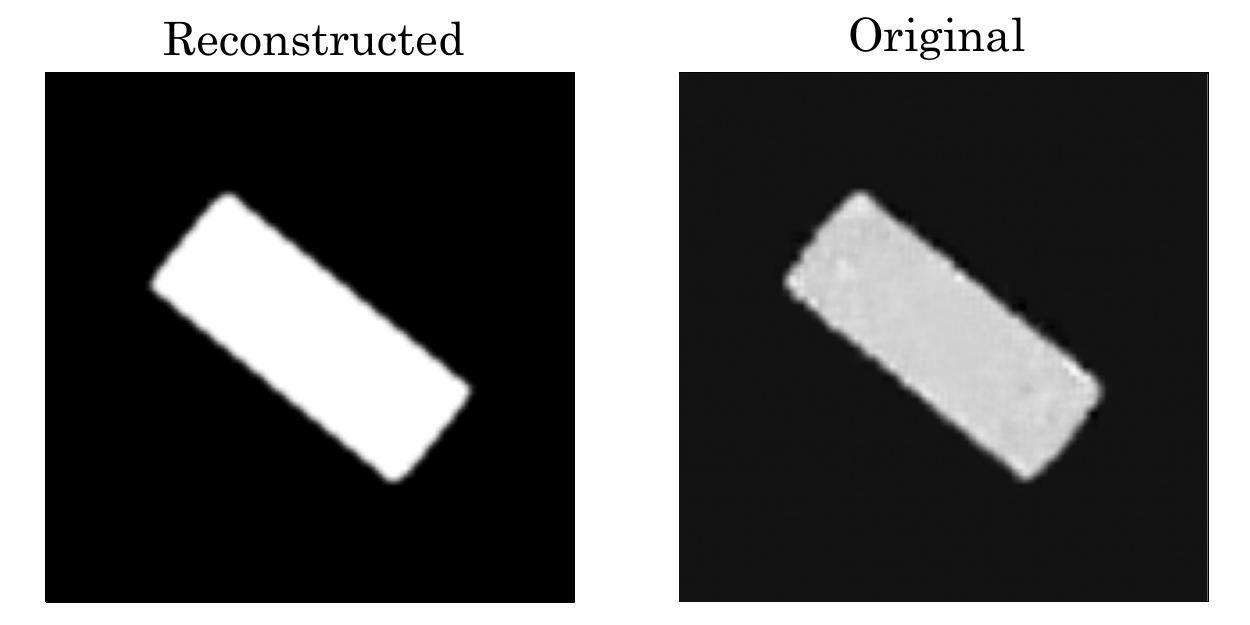}}
	\
	\subfloat[Example from \textbf{test} dataset.]{\label{RRAEs_Geometry:TvsR:Test1}\includegraphics[width=0.45\textwidth]{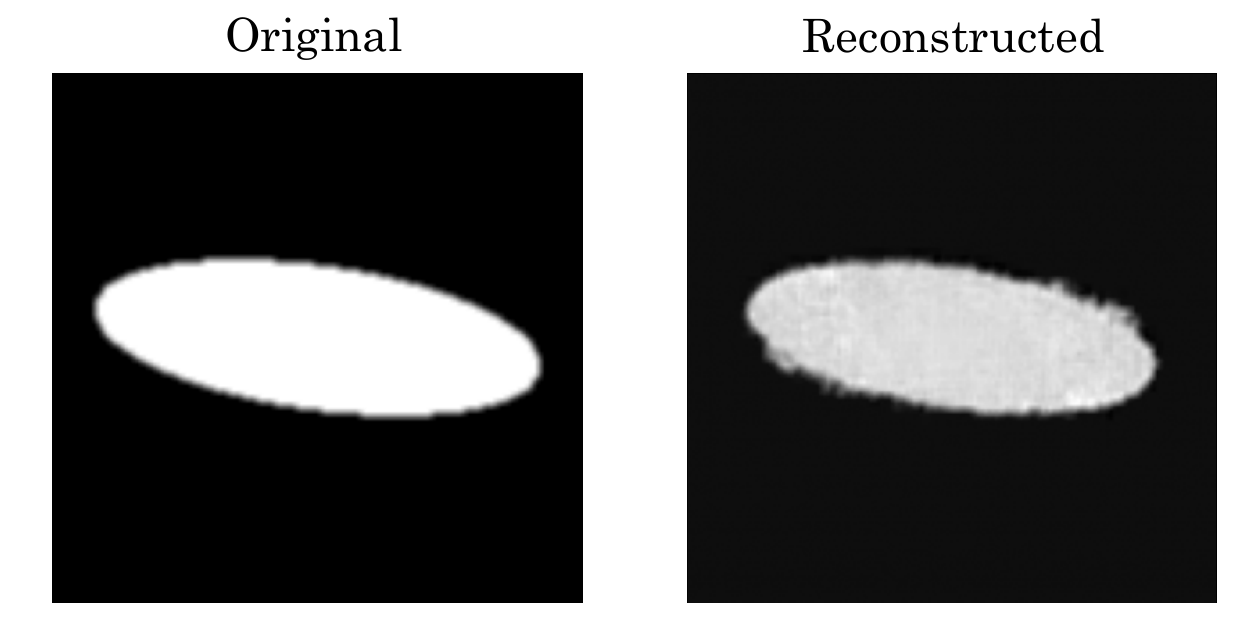}}
	\caption{Comparison of true and reconstructed RVE fiber geometries using the RRAEs model on training and test sets.}
	\label{fig:RRAEs_Geometry:True_vs_Reconstructed}
\end{figure}

For a qualitative evaluation of the trained RRAEs model, we selected two microstructures, one from the training set and one from the test set, and compared their original and reconstructed geometries. Each original microstructural image was first passed through the encoder $\mathcal{E}$, producing a high-dimensional representation that was subsequently reduced via truncated SVD to its four-dimensional latent vector $\boldsymbol{\alpha}$. This is then projected onto the basis matrix $\boldsymbol{U}$ and passed through the decoder $\mathcal{D}$ to reconstruct the microstructure. These resulting original/reconstructed image pairs are presented side by side in Figure \ref{fig:RRAEs_Geometry:True_vs_Reconstructed}. Visual comparison reveals a strong agreement between the original and reconstructed microstructures, indicating the model’s effectiveness in capturing and reproducing essential geometric features for both seen and unseen microstructures.

\begin{figure}[hbt!]
	\centering	\includegraphics[width=0.95\textwidth]{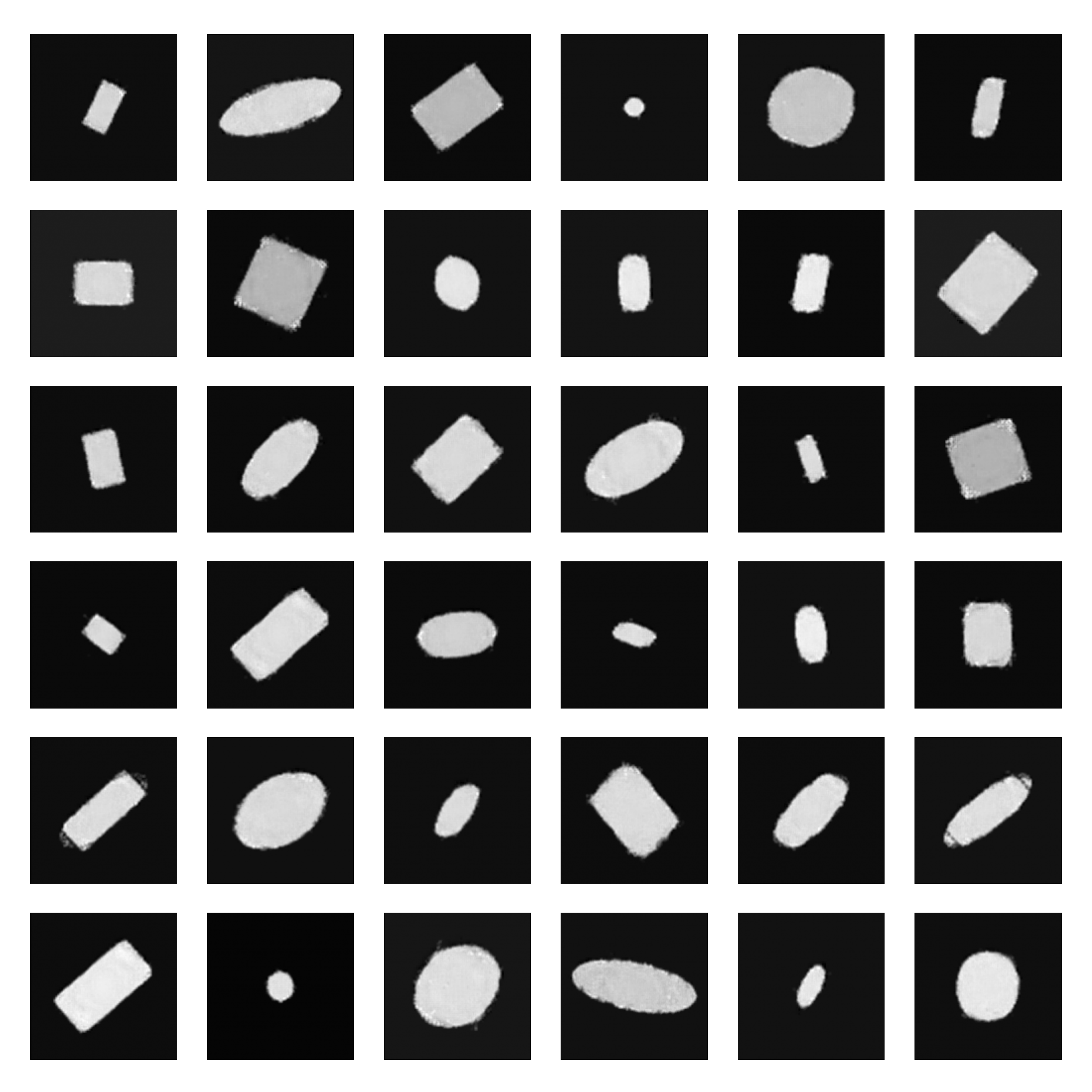}
	\caption{Newly generated microstructure: Obtained by sampling the latent-space coefficients $\boldsymbol{\alpha}$ using a Gaussian Mixture Model and decoding them to yield microstructure images.}
	\label{fig:RRAEs_Geometry:Genarative_Design}
\end{figure}

Figure \ref{fig:RRAEs_Geometry:Genarative_Design} shows a representative subset of 36 microstructures out of 2000 generated samples. These were obtained by sampling latent space coefficients $\boldsymbol{\alpha}$ using a Gaussian Mixture Model (GMM) \citep{bishop2006pattern}. The number of Gaussian components was selected by minimizing the Bayesian Information Criterion (BIC) over candidate values from 1 to 30. The GMM was trained on the same latent representations used for training the RRAEs, employing the Expectation Maximization (EM) algorithm with full covariance matrices and random initialization. Once the optimal model was identified, 2000 latent samples were generated and decoded through the trained RRAEs decoder $\mathcal{D}$ to reconstruct the corresponding microstructures.

To evaluate the generated microstructures, we computed the Fréchet Inception Distance (FID) \citep{heusel2017gans}. This metric measures the distance between the feature distributions of real and generated images extracted by a pre-trained Inception network; lower FID values indicate greater visual and statistical similarity. Using the original training set as reference, the FID between 2000 GMM-generated samples and the training set is 49, indicating a close match to the global characteristics of the training distribution. For a deeper interpretation of the FID scores, we further evaluated the test dataset, which consists of 99 microstructures, and compared it with an equal number of GMM-generated samples. The resulting FID values were 68 and 67, respectively. The near-identical FID values for the test set and the generated sample set indicate that the GMM-generated microstructures are approximately as similar to the training distribution as the held-out test images. Overall, these results highlight the ability of the proposed RRAEs framework to effectively explore the latent design space and generate morphologically realistic microstructures. Further improvements in FID could be achieved by increasing the training dataset, refining the RRAEs architecture or training procedure, and applying targeted post-processing to the generated images.





\FloatBarrier
\subsection{RRAEs for capturing the latent features of PGD Modes}
Once the multiparametric solution for predicting the von Mises stress $\sigma(\boldsymbol{x}, \mu_1, \mu_2)$ is constructed for every microstructure using the sPGD separated representation method, only the first three modes are retained. This choice provides an accurate, yet efficient approximation. However, if higher accuracy is required, additional modes can be included. In such cases, the number of modes may vary across microstructures; therefore, to maintain consistent structured data across all solutions, missing modes are padded with zeros. 

As explained previously in Section \ref{sec2:Generative Parametric Design} and illustrated in Figure \ref{fig:Generative_Parametric_Design}, the development of the GPD framework requires training RRAEs for the PGD separated functions. In this work, three RRAEs models are designed, each corresponding to one of the separated functions in the sPGD approximation: the spatial function $F(\boldsymbol{x})$, and the parametric functions $M_{1}(\mu_{1})$ and $M_{2}(\mu_{2})$, shown in Figures \ref{fig:RRAEs_GPD_Geometry:model}, \ref{fig:RRAEs_GPD_P1:model}, and \ref{fig:RRAEs_GPD_P2:model}, respectively. The first RRAEs model employs a Convolutional Neural Network (CNN)-based architecture to compress the spatial fields {$F^{1}(\boldsymbol{x})$, $F^{2}(\boldsymbol{x})$, $F^{3}(\boldsymbol{x})$} into a regularized latent space represented by the coefficients $\boldsymbol{\gamma}_{\boldsymbol{x}}$. The second and third RRAEs models, based on MultiLayer Perceptrons (MLPs), encode the one-directional parametric curves \{$M_{1}^{1}(\mu_1)$, $M_{1}^{2}(\mu_1)$, and $M_{1}^{3}(\mu_1)$\} and \{$M_{2}^{1}(\mu_2)$, $M_{2}^{2}(\mu_2)$, and $M_{2}^{3}(\mu_2)$\} into latent space coefficients $\boldsymbol{\gamma}_{1}$ and $\boldsymbol{\gamma}_{2}$, respectively. These latent representations provide a compact, continuous, and regularized description of both spatial and parametric variations, enabling efficient reconstruction and interpolation of the multiparametric solution within a reduced basis.

\subsubsection{RRAEs representation of the spatial PGD functions $F^{i}$}
\begin{figure}[hbt!]
	\centering
	\includegraphics[width=0.95\textwidth]{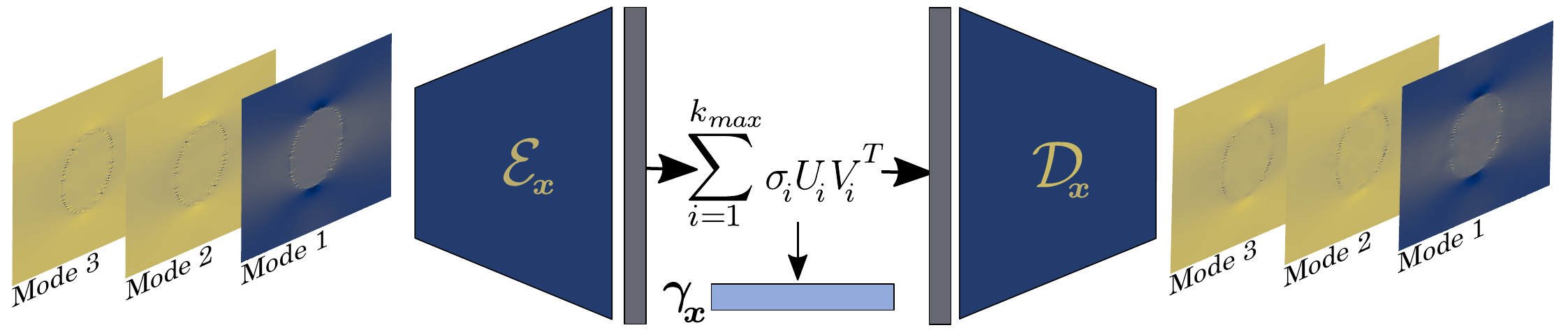}
	\caption{RRAEs model for representing the spatial modes $F^{i}(\boldsymbol{x})$ of the sPGD approximation.}
	\label{fig:RRAEs_GPD_Geometry:model}
\end{figure}

Given that the spatial function $F(\boldsymbol{x})=F(x,y)$, is defined over two dimensions, an image-based representation is used to train the RRAE model. Specifically, the input data are formatted as images of size $148\times148\times3$, where the first two values correspond to the spatial resolution (pixels along $x$ and $y$ directions), and the last value represents the number of channels, corresponding here to the number of PGD modes. This image-like structure allows the convolutional RRAEs to effectively capture spatial correlations and mode interactions across the microstructure.

As shown in Figure \ref{fig:RRAEs_GPD_Geometry:model} and explained in Section \ref{subsec:RRAEs}, the input data are passed through the encoder $\mathcal{E}_{\boldsymbol{x}}$ into a latent space of size $L=3000$. Within this space, a truncated SVD is applied, retaining $k_{\max}=12$ singular values. This produces the latent coefficient vector $\boldsymbol{\gamma}_{\boldsymbol{x}} = [\gamma_{\boldsymbol{x}1}, \ldots, \gamma_{\boldsymbol{x}12}]$. Then the Decoder $\mathcal{D}_{\boldsymbol{x}}$ reconstructs the original PGD modes from the latent coefficients $\boldsymbol{\gamma}_{\boldsymbol{x}}$ after projection with the corresponding truncated SVD matrix $\mathbf{U}^{\boldsymbol{x}}$. The training is carried out using the AdaBelief optimizer with a batch size of 20. A staged learning rate schedule is adopted, consisting of three phases of 6000, 5000, 4000 steps, with learning rates set to $10^{-3}$, $10^{-4}$, and $10^{-5}$, respectively. Table \ref{tab:RRAEs_hyperparameters} lists the main hyperparameters used for training, including activation functions, the number and size of layers in the convolutional and MLP branches, and convolutional settings (kernel size, stride and padding).

\begin{figure}[bht!]
	\centering              
	\subfloat[\scriptsize{MAE on the \textbf{training} dataset for each mode.}]{\label{subfig:MAE_training_x}\includegraphics[width=0.45\textwidth]{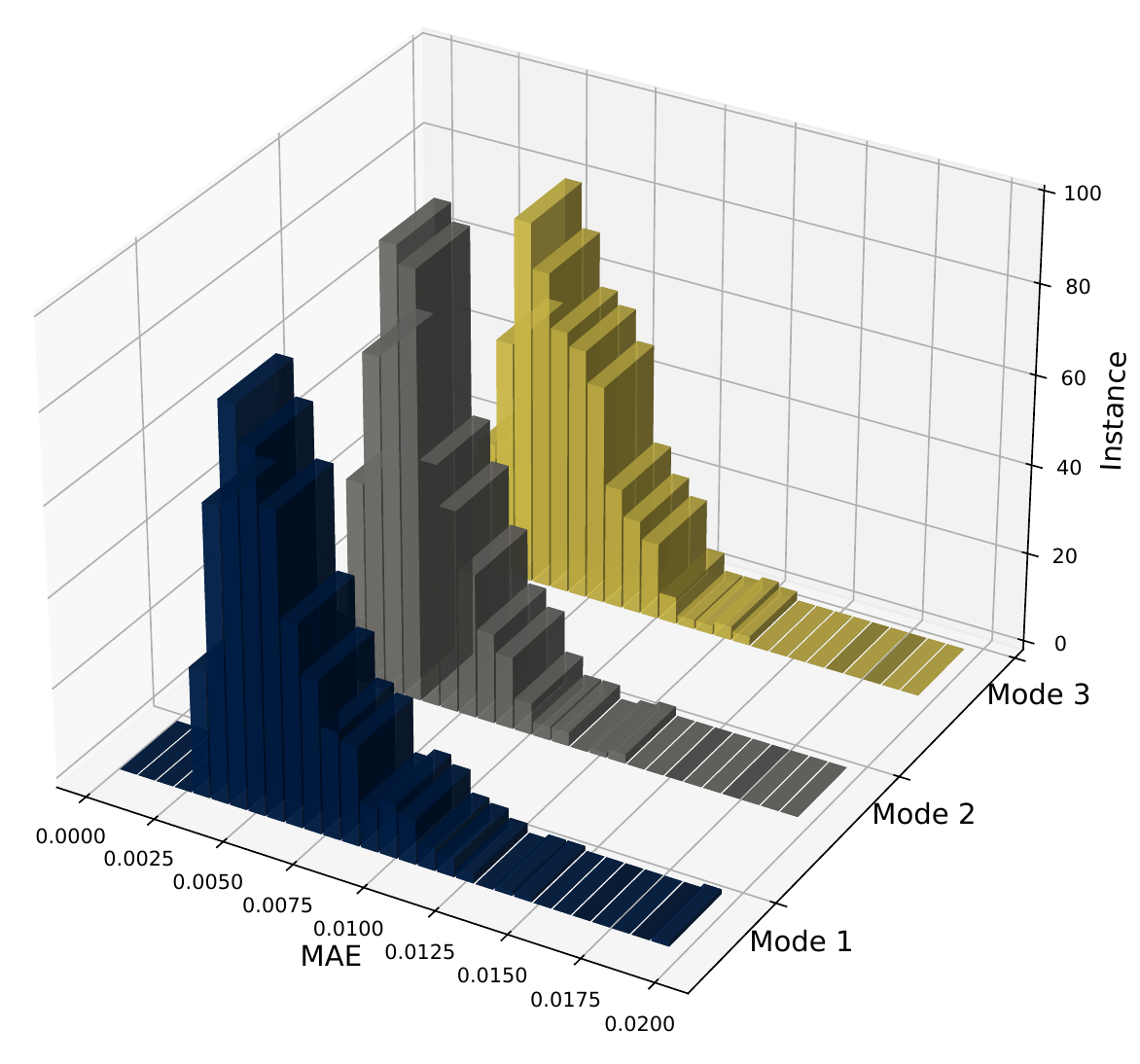}}
	\
	\subfloat[\scriptsize{MAE on the \textbf{test} dataset for each mode.}]{\label{subfig:MAE_test_x}\includegraphics[width=0.45\textwidth]{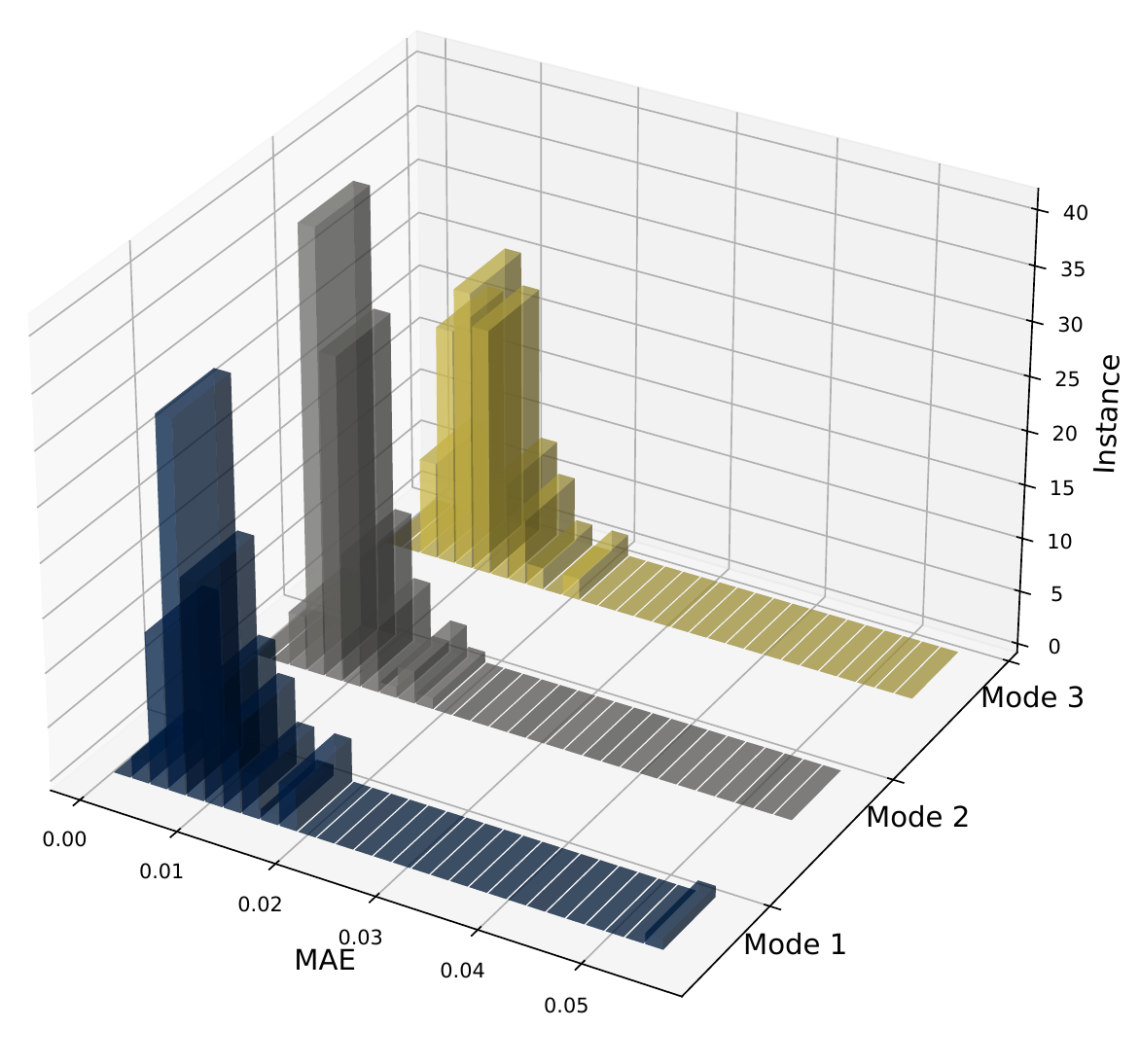}}
	\
	\subfloat[\scriptsize{Original vs. Reconstructed spatial function $F^{i}(\boldsymbol{x})$ for each mode $i$, on a \textbf{training} microstructure.}]{\label{subfig:Comparison_training_x}\includegraphics[width=0.45\textwidth]{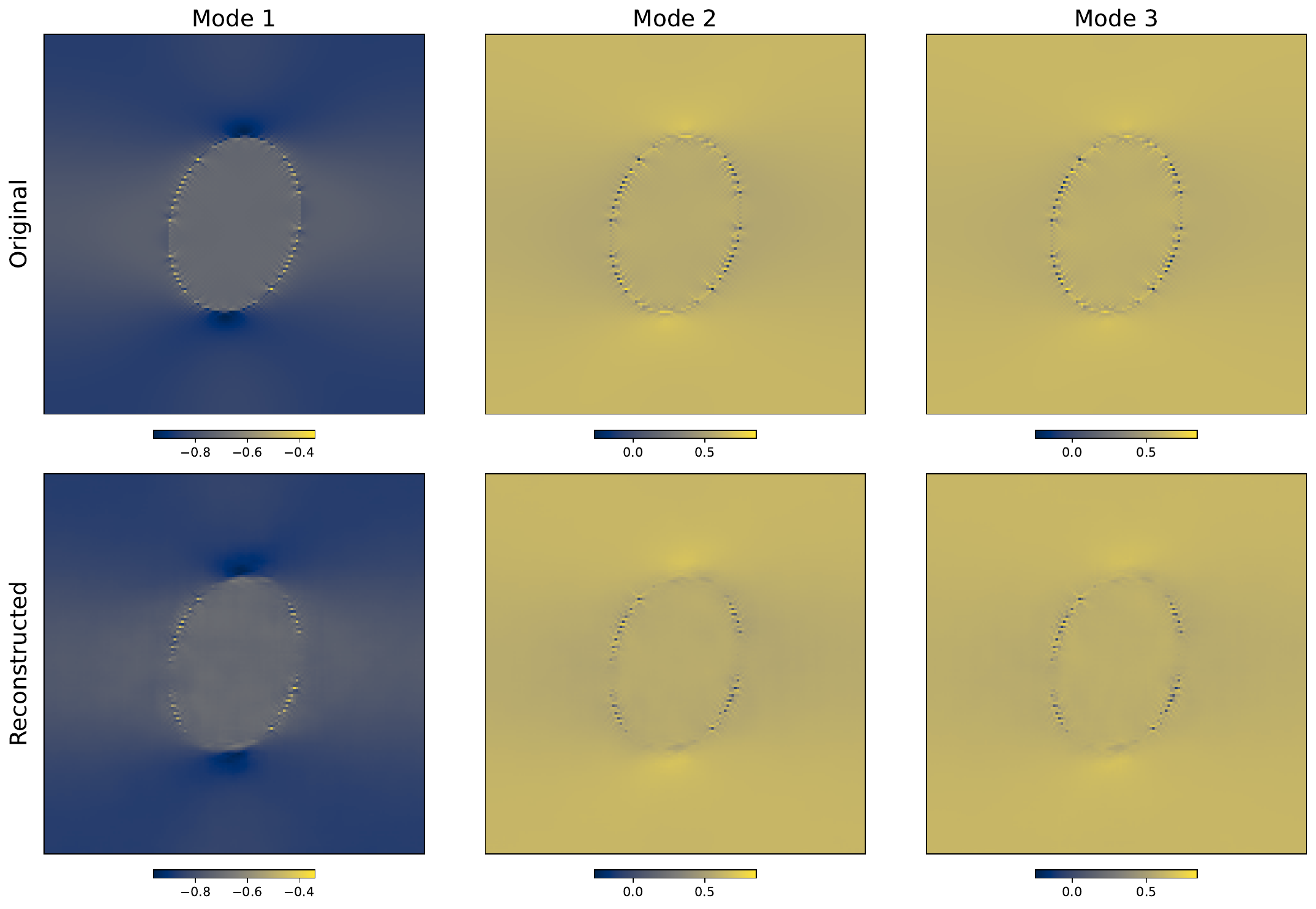}}
	\
	\subfloat[\scriptsize{Original vs. Reconstructed spatial function $F^{i}(\boldsymbol{x})$ for each mode $i$, on a \textbf{test} microstructure.}]{\label{subfig:Comparison_test_x}\includegraphics[width=0.45\textwidth]{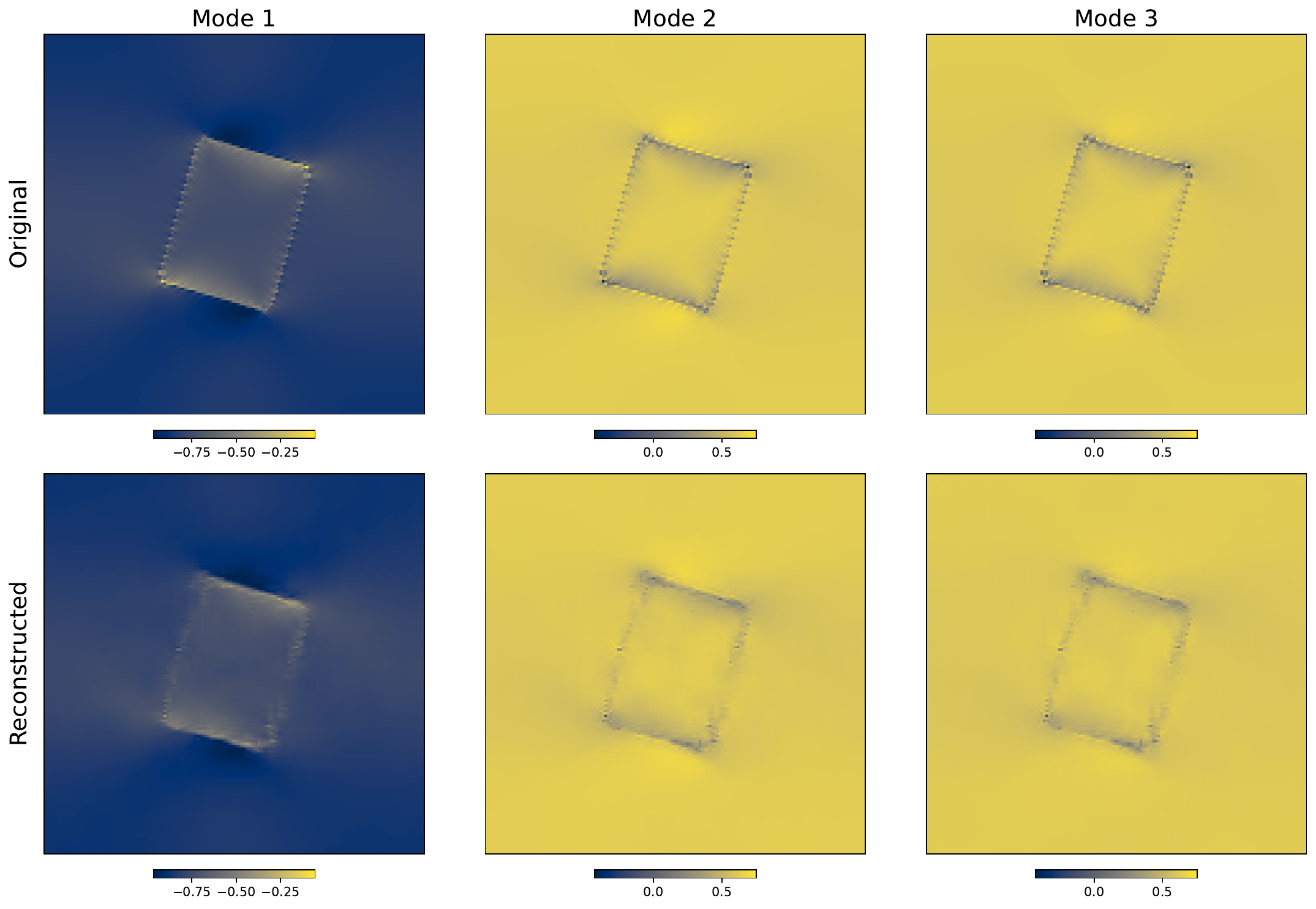}}
	\caption{Evaluation of the trained RRAEs model for the sPGD separated representation modes $F^{i}(\boldsymbol{x})$: (a) and (b) show the MAE between the original curves from sPGD and those reconstructed by the RRAEs for each mode $i$, computed on the training and test datasets, respectively. (c) and (d) compare the normalized values of original and reconstructed curves for various modes on two randomly selected microstructures, one from the training set and one from the test set, respectively.}
	\label{fig:RRAEs_evaluation_x}
\end{figure}

The trained RRAEs model is evaluated in two steps. First, the Mean Absolute Error (MAE) is computed for each mode on the training and test sets between the original and reconstructed images; the resulting histograms are shown in Figures \ref{subfig:MAE_training_x} and \ref{subfig:MAE_test_x}. The consistently low errors indicate that the model accurately reconstructs the separated sPGD function modes. Second, Figures \ref{subfig:Comparison_training_x} and \ref{subfig:Comparison_test_x} present a visual comparison for one sample from the training set and one from the test set. Each figure shows the three normalized spatial modes from the original data (top row) alongside the corresponding RRAEs reconstructions (bottom row) (Figure \ref{fig:RRAEs_GPD_Geometry:model}). The reconstructions closely match the originals, preserving the key spatial patterns. While minor smoothing of fine details occurs, the essential features are retained, confirming that the RRAEs provide reliable mode reconstruction. Importantly, the modes are ranked by significance according to the sPGD model: Mode 1 captures the dominant large-scale structure, whereas Modes 2 and 3 represent progressively finer variations.

\FloatBarrier
\subsubsection{RRAEs representation of the parametric PGD functions $M_{1}^{i}$}
\begin{figure}[hbt!]
	\centering
	\includegraphics[width=0.95\textwidth]{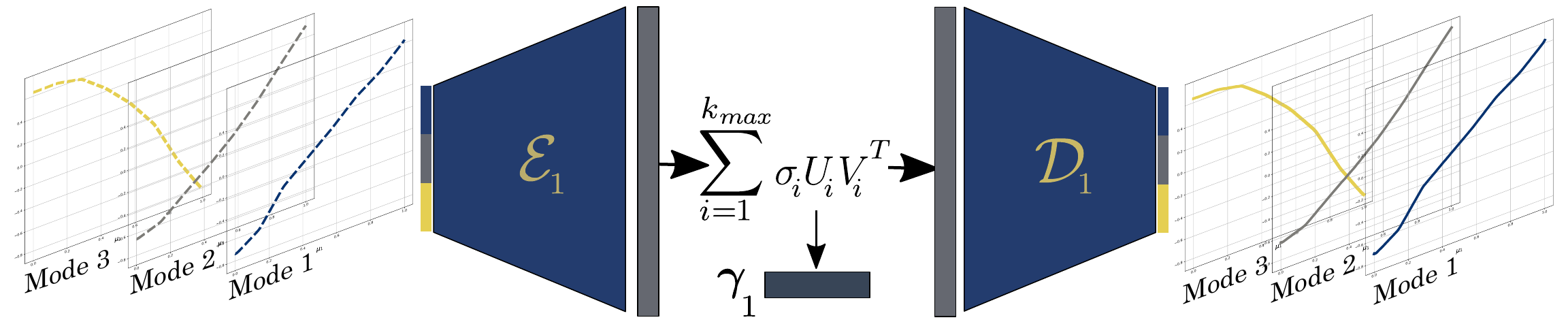}
	\caption{RRAEs model for representing the unidirectional parametric modes $M_{1}^{i}(\mu_{1})$ (curves) in the sPGD approximation.}
	\label{fig:RRAEs_GPD_P1:model}
\end{figure}

Following the same methodology as in the previous subsection, RRAEs model is employed to project the parametric modes of the separated PGD approximation ($M_{1}^{i}$) into a compact and regularized latent space represented by the coefficients $\boldsymbol{\gamma}_{1}$, as illustrated in Figure \ref{fig:RRAEs_GPD_P1:model}. Unlike the spatial case, where a CNN-based RRAE was used, here an MLP-based RRAE is adopted, as the input and output are one-dimensional vectors rather than images. Each curve $M_{1}^{i}(\mu_{1})$ is initially sampled at 1000 points along the $\mu_{1}$ variable and normalized to the range [-1, 1]. Considering the first three modes, the data are reshaped into a single vector of 3000 elements. The input vector is passed through the encoder $\mathcal{E}_{1}$ into a latent space of size $L=1700$. In latent space a truncated SVD is applied and reduced to $k_{\max}=3$ leading to a compact coefficient vector $\boldsymbol{\gamma}_{1}\in \mathbb{R}^{k_{\max}}$. Reconstruction is performed by projecting $\boldsymbol{\gamma}_{1}$ onto the truncated basis $\mathbf{U}^{1}$ and passing the result to the decoder $\mathcal{D}{1}$, which recovers the original curves.

Training for this MLP-based RRAE is performed with the AdaBelief optimizer and a batch size of 20. A staged learning-rate schedule with three phases ( 3000 steps each) is used, with learning rates $10^{-3}$, $10^{-4}$, and $10^{-5}$ to promote stable convergence and progressive fine-tuning. Key hyperparameters, including activation functions, the number and size of MLP layers, and other architectural settings, are summarized in Table~\ref{tab:RRAEs_hyperparameters}.

\begin{figure}[bht!]
	\centering              
	\subfloat[\scriptsize{MAE on the \textbf{training} dataset for each mode.}]{\label{subfig:MAE_training_P1}\includegraphics[width=0.45\textwidth]{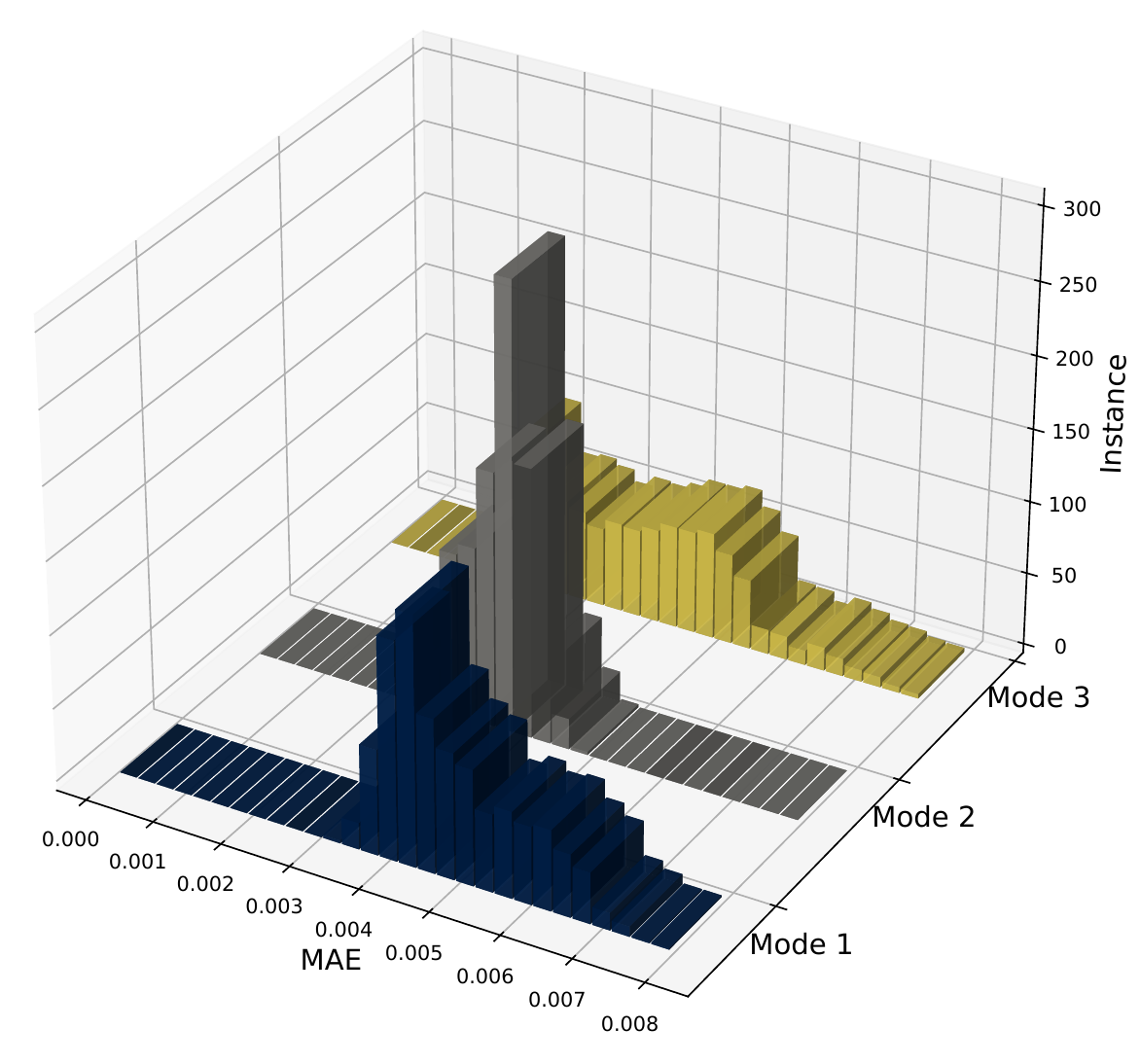}}
	\
	\subfloat[\scriptsize{MAE on the \textbf{test} dataset for each mode.}]{\label{subfig:MAE_test_P1}\includegraphics[width=0.45\textwidth]{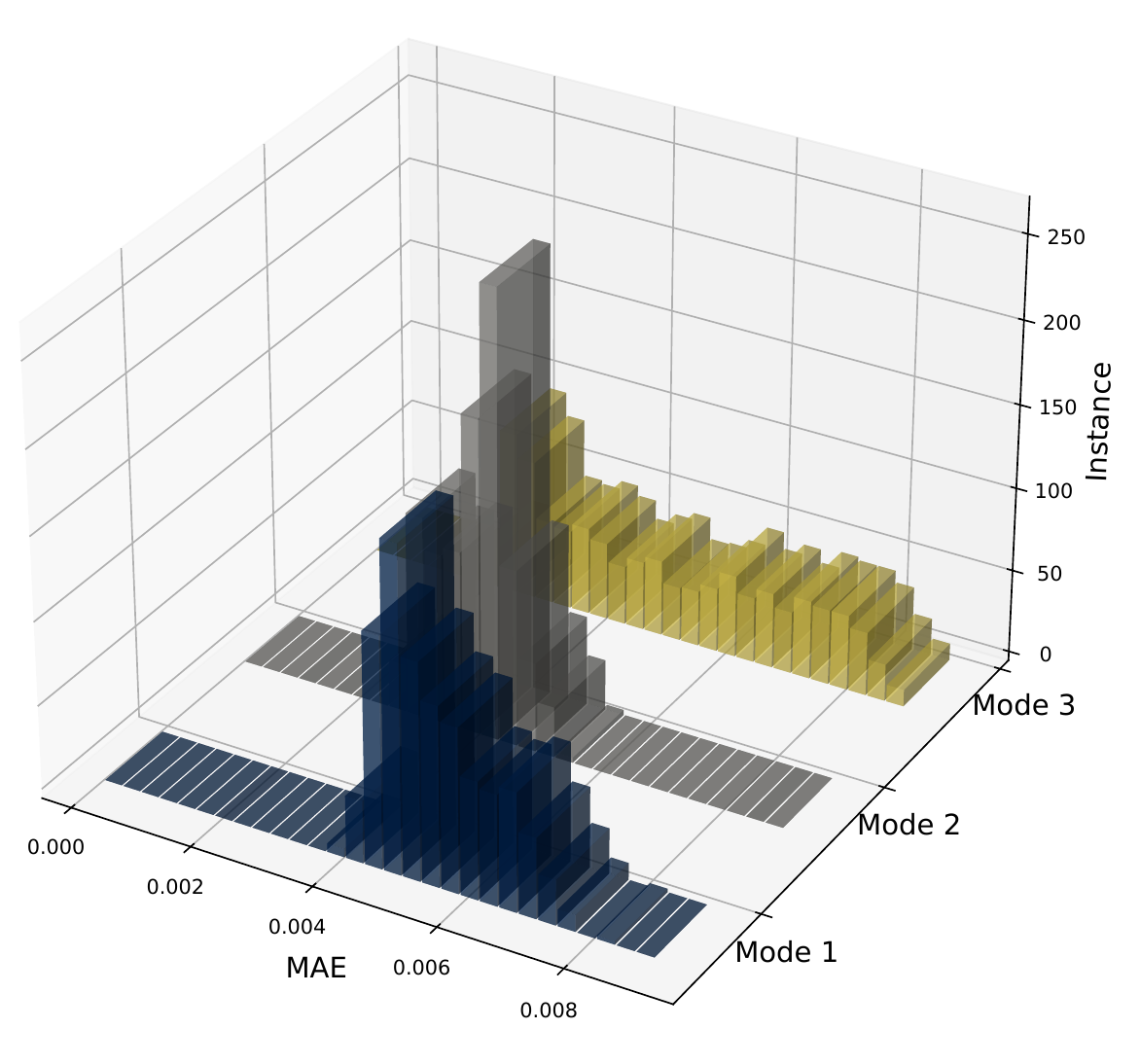}}
	\
	\subfloat[\scriptsize{Original vs. Reconstructed curves $M^{i}_{1}(\mu_{1})$ on a \textbf{training} microstructure for each mode $i$.}]{\label{subfig:Comparison_training_P1}\includegraphics[width=0.45\textwidth]{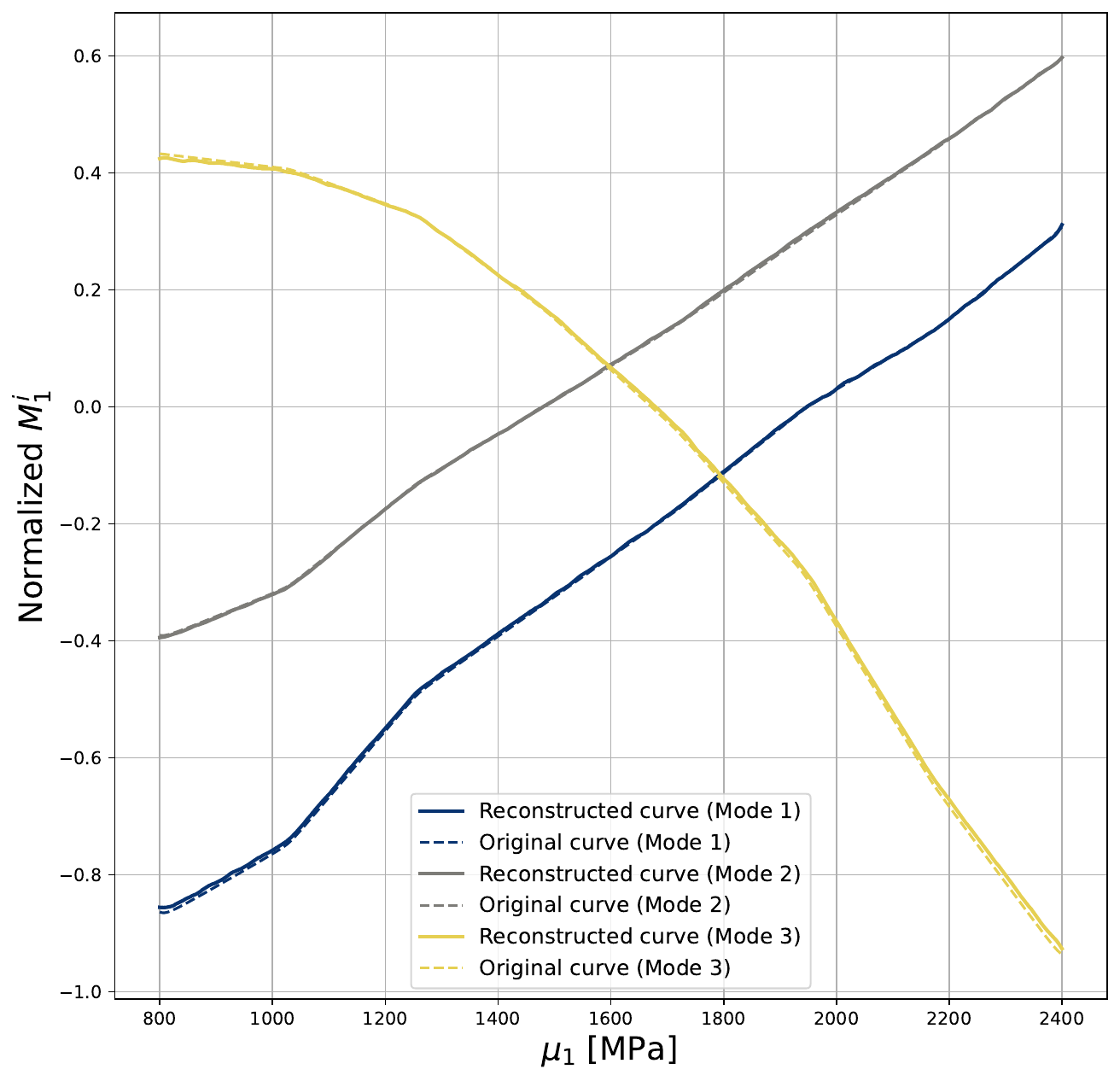}}
	\
	\subfloat[\scriptsize{Original vs. Reconstructed curves $M^{i}_{1}(\mu_{1})$ on a \textbf{test} microstructure for each mode $i$.}]{\label{subfig:Comparison_test_P1}\includegraphics[width=0.45\textwidth]{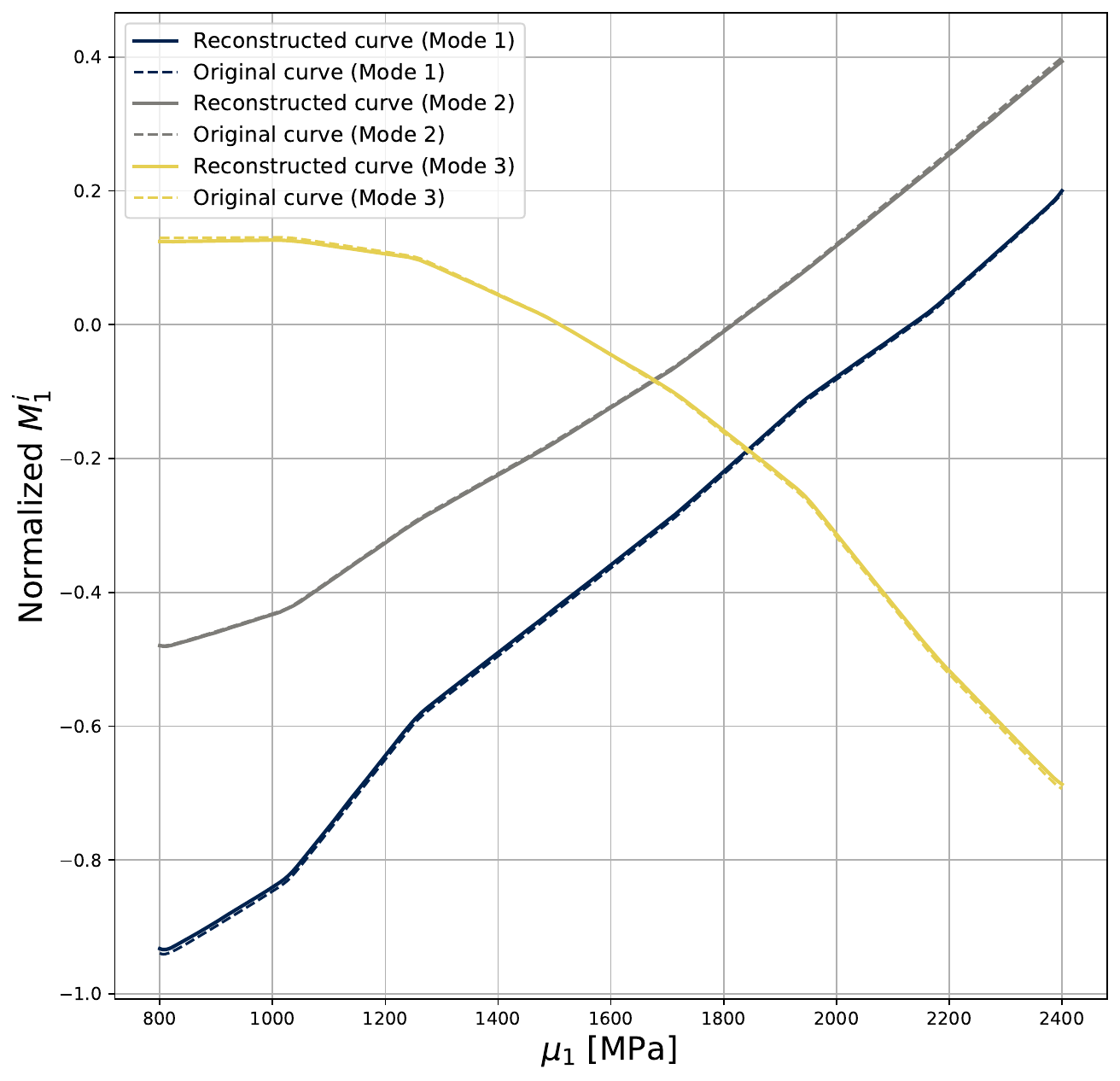}}
	\caption{Evaluation of the trained RRAEs model for the sPGD separated representation modes $M_{1}^{i}(\mu_{1})$: (a) and (b) show the MAE between the original curves from sPGD and those reconstructed by the RRAEs for each mode $i$, computed on the training and test datasets, respectively. (c) and (d) compare the normalized values of original and reconstructed curves for various modes on two randomly selected microstructures, one from the training set and one from the test set, respectively.}
	\label{fig:RRAEs_evaluation_P1}
\end{figure}

Figures \ref{subfig:MAE_training_P1} and \ref{subfig:MAE_test_P1} report the MAE between the original and reconstructed curves for the three sPGD modes, evaluated on both the training and test sets. In both datasets the MAE values are consistently low and concentrated mostly below 0.007, demonstrating high reconstruction accuracy. The training and test error distributions are very similar across all modes, providing no evidence of overfitting and indicating good generalization.

The accompanying subplots compare the original and reconstructed curves $M_{1}^{i}(\mu_{1})$ for modes $i=1,2,3$ as functions of the parameter $\mu_{1}$: panel (\ref{subfig:Comparison_training_P1}) shows a training microstructure and panel (\ref{subfig:Comparison_test_P1}) a test microstructure. In both cases the reconstructions closely follow the original trends for each mode. These results confirm that the trained RRAEs model compresses the three curves (3000 points) into just three latent coefficients, allowing accurate reconstruction. This compact representation simplifies and speeds up any regression from the design parameters $\boldsymbol{\alpha}$ to the modal features, without losing essential information.

\FloatBarrier
\subsubsection{RRAEs representation of the parametric PGD functions $M_{2}^{i}$}
\begin{figure}[hbt!]
	\centering
	\includegraphics[width=0.95\textwidth]{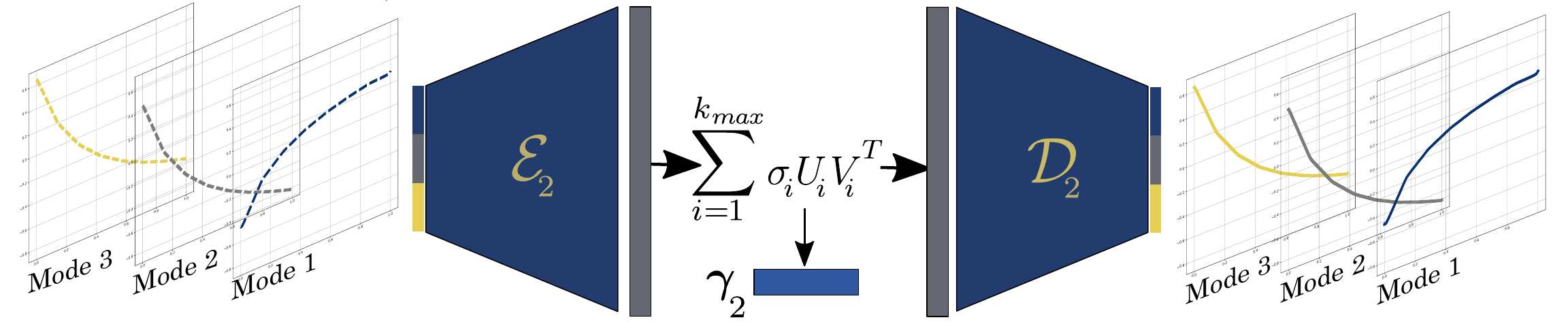}
	\caption{RRAEs model for representing the unidirectional parametric modes $M_{2}^{i}(\mu_{2})$ (curves) in the sPGD approximation}
	\label{fig:RRAEs_GPD_P2:model}
\end{figure}

As illustrated in Figure \ref{fig:RRAEs_GPD_P2:model}, here we also used the same architecture of RRAEs model as in the previous subsection; RRAEs model is employed to project high-dimensional separated PGD functions modes into compact and regularized latent space coefficients $\boldsymbol{\gamma}_{2}$. This dimensionality reduction facilitates efficient manipulation, storage, and learning of the underlying structure of the solution space. Each mode $M_{2}^{i}(\mu_{2})$, originally sampled at 1000 spatial points, is first normalized to the range [-1,1], and then concatenated into a single 3000-element vector for input to the unsupervised RRAEs model. The training procedure follows the same hyperparameters as for $M_{1}^{i}$, except that the latent space dimension is set to $L=800$, and the SVD truncation is fixed at $k_{\max}=3$.
\begin{figure}[bht!]
	\centering              
	\subfloat[\scriptsize{MAE on the \textbf{training} dataset for each mode.}]{\label{subfig:MAE_training_P2}\includegraphics[width=0.45\textwidth]{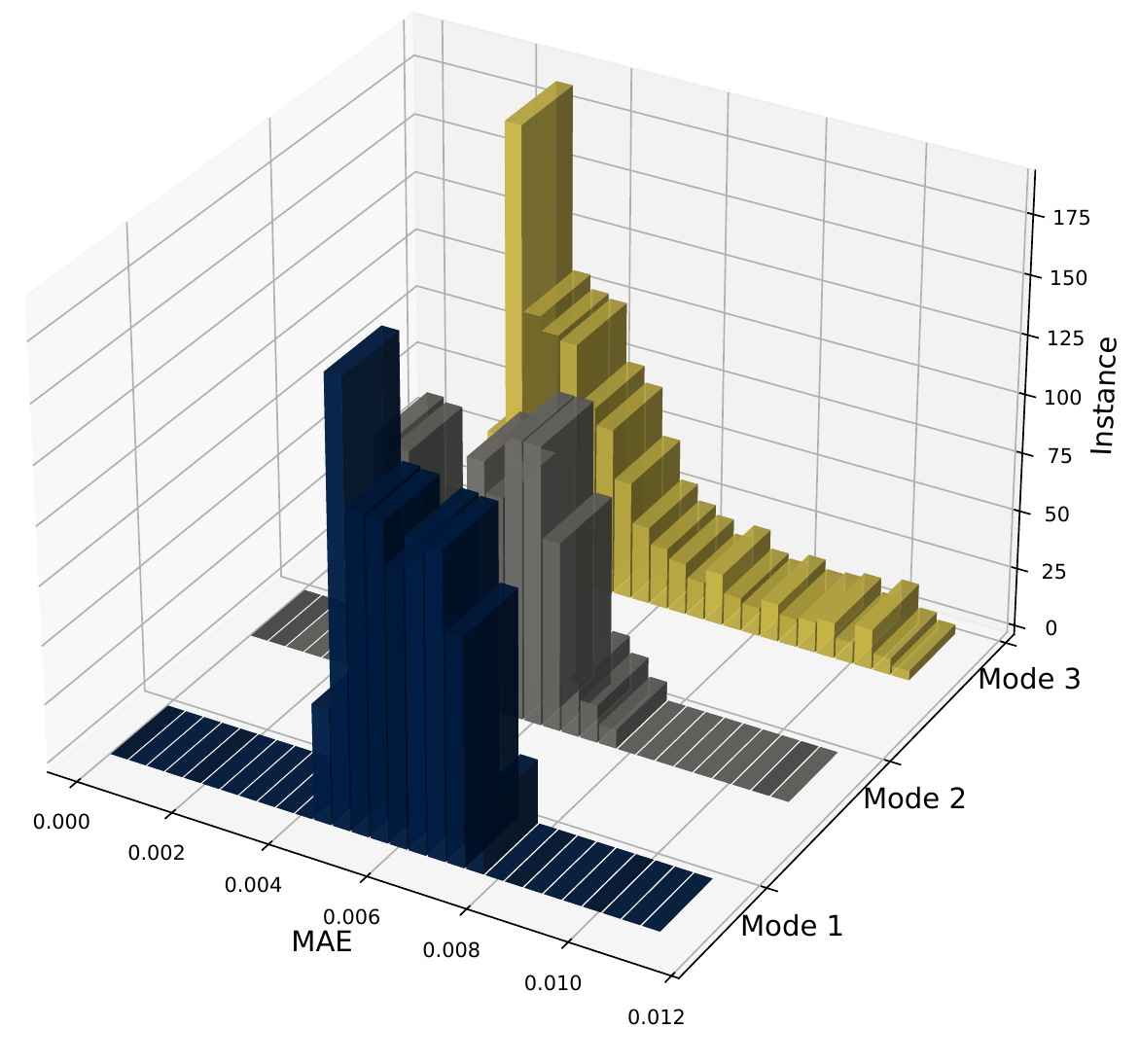}}
	\
	\subfloat[\scriptsize{MAE on the \textbf{test} dataset for each mode.}]{\label{subfig:MAE_test_P2}\includegraphics[width=0.45\textwidth]{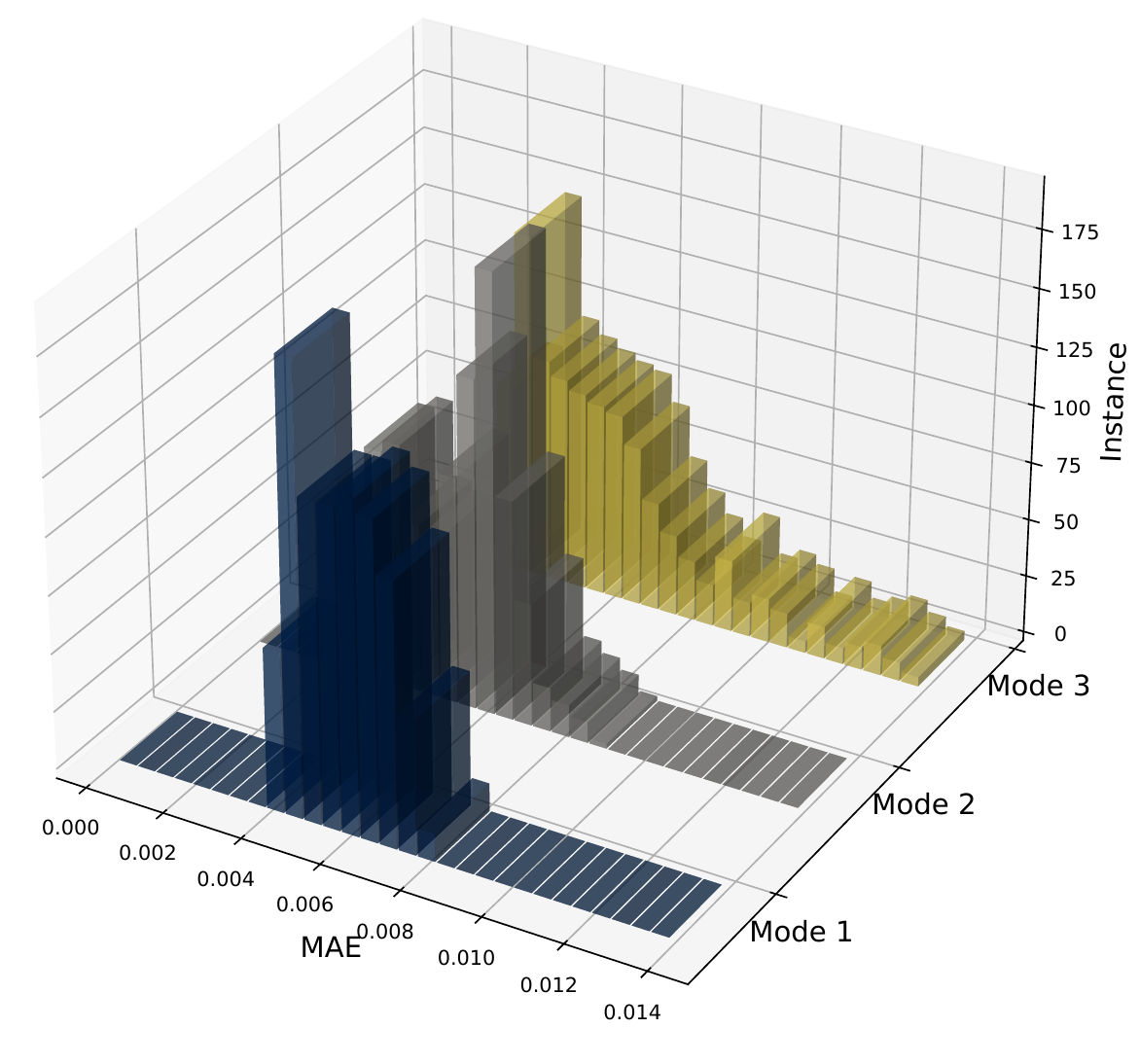}}
	\
	\subfloat[\scriptsize{Original vs. Reconstructed curves $M^{i}_{2}(\mu_{2})$ on a \textbf{training} microstructure for each mode $i$.}]
	{\label{subfig:Comparison_training_P2}\includegraphics[width=0.45\textwidth]{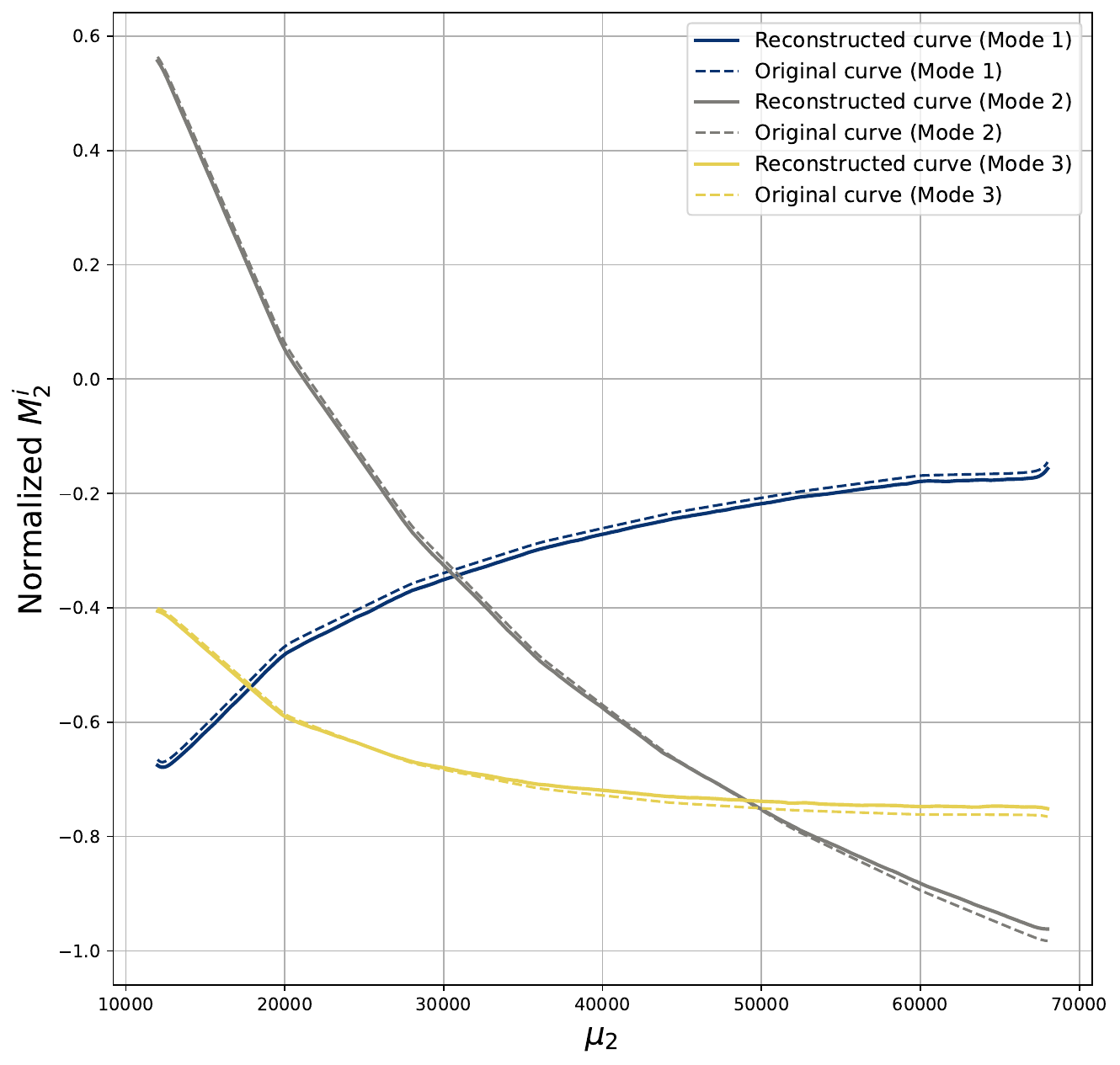}}
	\
	\subfloat[\scriptsize{Original vs. Reconstructed curves $M^{i}_{2}(\mu_{2})$ on a \textbf{test} microstructure for each mode $i$.}]{\label{subfig:Comparison_test_P2}\includegraphics[width=0.45\textwidth]{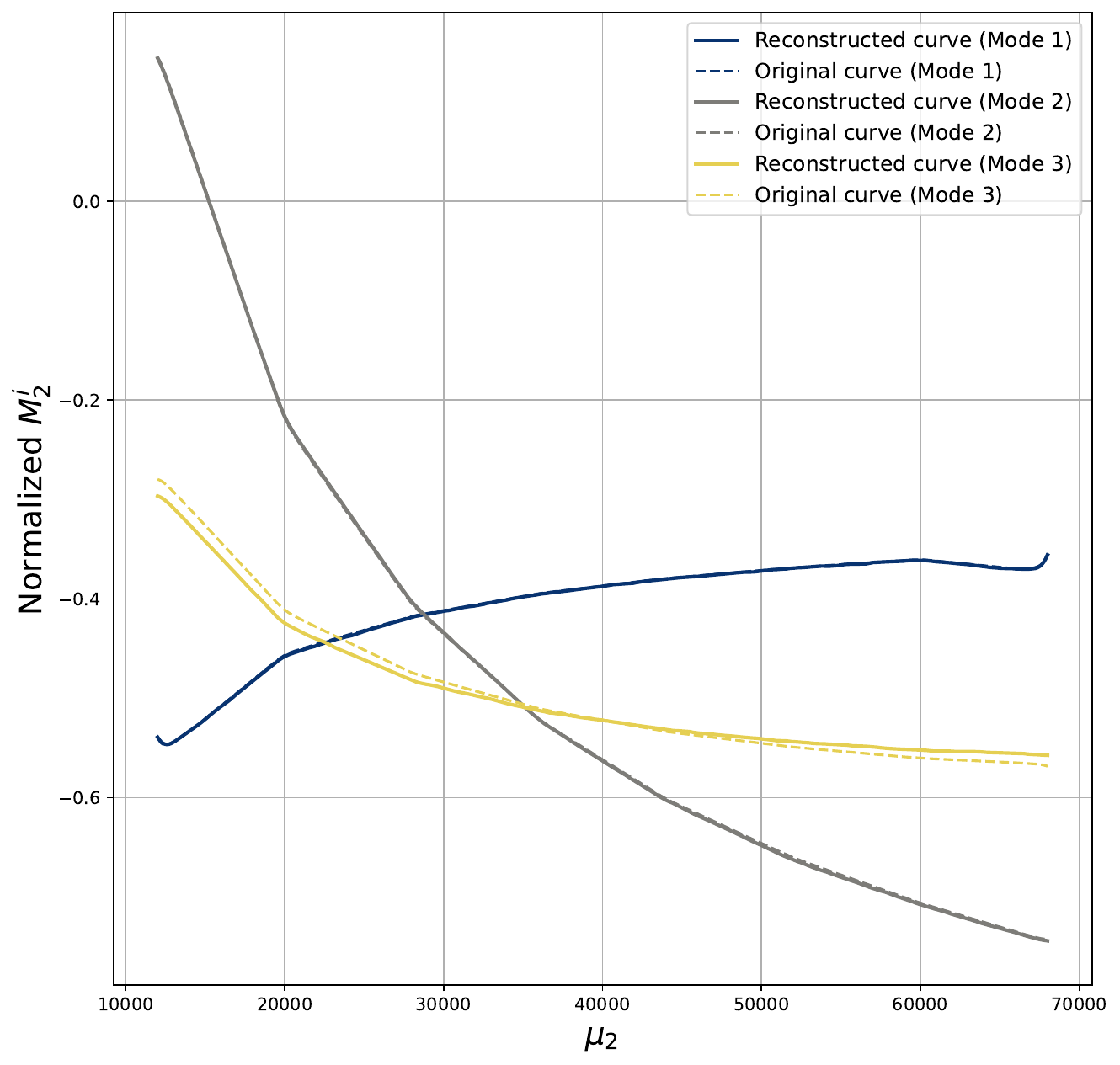}}
	\caption{Evaluation of the trained RRAEs model for the sPGD separated representation modes $M_{2}^{i}(\mu_{2})$: (a) and (b) show the MAE values between the original curves from sPGD and those reconstructed by the RRAEs for each mode $i$, computed on the training and test datasets, respectively. (c) and (d) compare the normalized values of original and reconstructed curves for various modes on two randomly selected microstructures, one from the training set and one from the test set, respectively.}
	\label{fig:RRAEs_evaluation_P2}
\end{figure}

Figure \ref{fig:RRAEs_evaluation_P2} shows the distributions of the MAE between reference and reconstructed curves in the $\mu_2$ direction, for both training (Panel \ref{subfig:MAE_training_P2}) and test (Panel \ref{subfig:MAE_test_P2}) datasets,  for each of the three PGD modes on the training dataset. Each mode is visualized separately along the $z$-axis, with MAE values on the $x$-axis and their frequency on the $y$-axis. MAE values are generally low, mostly below 0.012, indicating that the RRAEs model accurately reproduces the modal contributions. For the training set, Modes 1 and 2 exhibit the sharpest peak at very small errors, while Mode 3 has broader distributions, with Mode 3 showing a longer tail toward higher errors. The test set displays similar trends, with slightly wider distributions, confirming the model’s generalization capability.

Figure \ref{fig:RRAEs_evaluation_P2} also compares original and reconstructed curves for a representative training microstructure (Panel \ref{subfig:Comparison_training_P2}) and a test microstructure (Panel \ref{subfig:Comparison_test_P2}). The reconstructed curves closely follow the originals, with Modes 1 and 2 nearly overlapping and Mode 3 showing minor deviations, especially in the test case, This suggests that the third mode, typically associated with more complex or subtle features, is more challenging to approximate accurately. Overall, these results demonstrate that the RRAEs model successfully captures the essential variation of the three modal curves in the $\mu_2$ direction using only three latent parameters, while maintaining strong generalization to unseen data.

\begin{table}[bht!]
	\scriptsize
	\begin{center}
		\caption {Hyperparameters of the RRAEs models for geometry and PGD modes.}
		\label{tab:RRAEs_hyperparameters}
		\begin{tabular}{l l c c c c c} 
			\hline
			RRAE &  & Hidden layers & \makecell{Activation \\ function} & Kernel size & Stride & Padding\\
			\hline
			\hline
			$\mathcal{E}$ & Convolution layers &  [32,64,128,256]  & ReLU & 4 & 2 & 1\\
			& MLP layers & [64,64] & Softplus &  &  & \\
			\cline{2-7}
			$\mathcal{D}$	& MLP layers & [64,64] & Softplus  &  &   &  \\
			&Convolution transpose layers &  [8,1] & ReLU & 4 & 2 & 1 \\
			&Convolution layers &  [1] & Linear & 1 & 1 & 0 \\
			\hline
			\hline
			$\mathcal{E}_{\boldsymbol{x}}$ & Convolution layers &  [32,64,128,256]  & ReLU & 3 & 2 & 1\\
			& MLP layers & [64,64] & Softplus &  &  & \\
			\cline{2-7}
			$\mathcal{D}_{\boldsymbol{x}}$	& MLP layers & [64,64] & Softplus  &  &   &  \\
			&Convolution transpose layers &  [64,32] & ReLU & 3 & 2 & 1 \\
			&Convolution layers &  [1] & Linear & 1 & 1 & 0 \\
			\hline
			\hline
			$\mathcal{E}_{1}$  & MLP layers & [64] & ReLU &  &  & \\
			\cline{2-7}
			$\mathcal{D}_{1}$	& MLP layers & [64,64,64,64,64,64] & ReLU  &  &   &  \\
			\hline
			\hline
			$\mathcal{E}_{2}$	& MLP layers & [64] & ReLU &  &  & \\
			\cline{2-7}
			$\mathcal{D}_{2}$	& MLP layers & [64,64,64,64,64,64] & ReLU  &  &   &  \\
			\hline
		\end{tabular}
	\end{center}
	\normalsize
\end{table}

\FloatBarrier
\subsection{From microstructure morphology (Geometry) to multiparametric solution (sPGD modes)}
After training the four RRAEs models for geometry and PGD modes, the latent coefficients of each latent space, denoted as $\boldsymbol{\alpha}$, $\boldsymbol{\gamma}_{\boldsymbol{x}}$, $\boldsymbol{\gamma}_{1}$, and $\boldsymbol{\gamma}_{2}$, are computed for both the training and test microstructure datasets through their respective encoders $\mathcal{E}$, $\mathcal{E}_{\boldsymbol{x}}$, $\mathcal{E}_{1}$, and $\mathcal{E}_{2}$, following the truncated SVD representation. The latent coefficients are then normalized, and a regression mapping $f$ is established to relate $\boldsymbol{\alpha}$ to the set [$\boldsymbol{\gamma}_{\boldsymbol{x}}$, $\boldsymbol{\gamma}_{1}$, $\boldsymbol{\gamma}_{2}$]. To this end, three independent MLP models are constructed. Each MLP employs ReLU activation functions in the hidden layers, the ADAM optimizer, a Mean Absolute Error loss function, and a batch size of 32. The first MLP consists of three hidden layers with 128 neurons each and is trained for 3000 epochs, while the second and third MLPs each consist of two hidden layers with 64 neurons and are trained for 2000 epochs.

\begin{figure}[hbt!]
	\centering
	\includegraphics[width=0.95\textwidth]{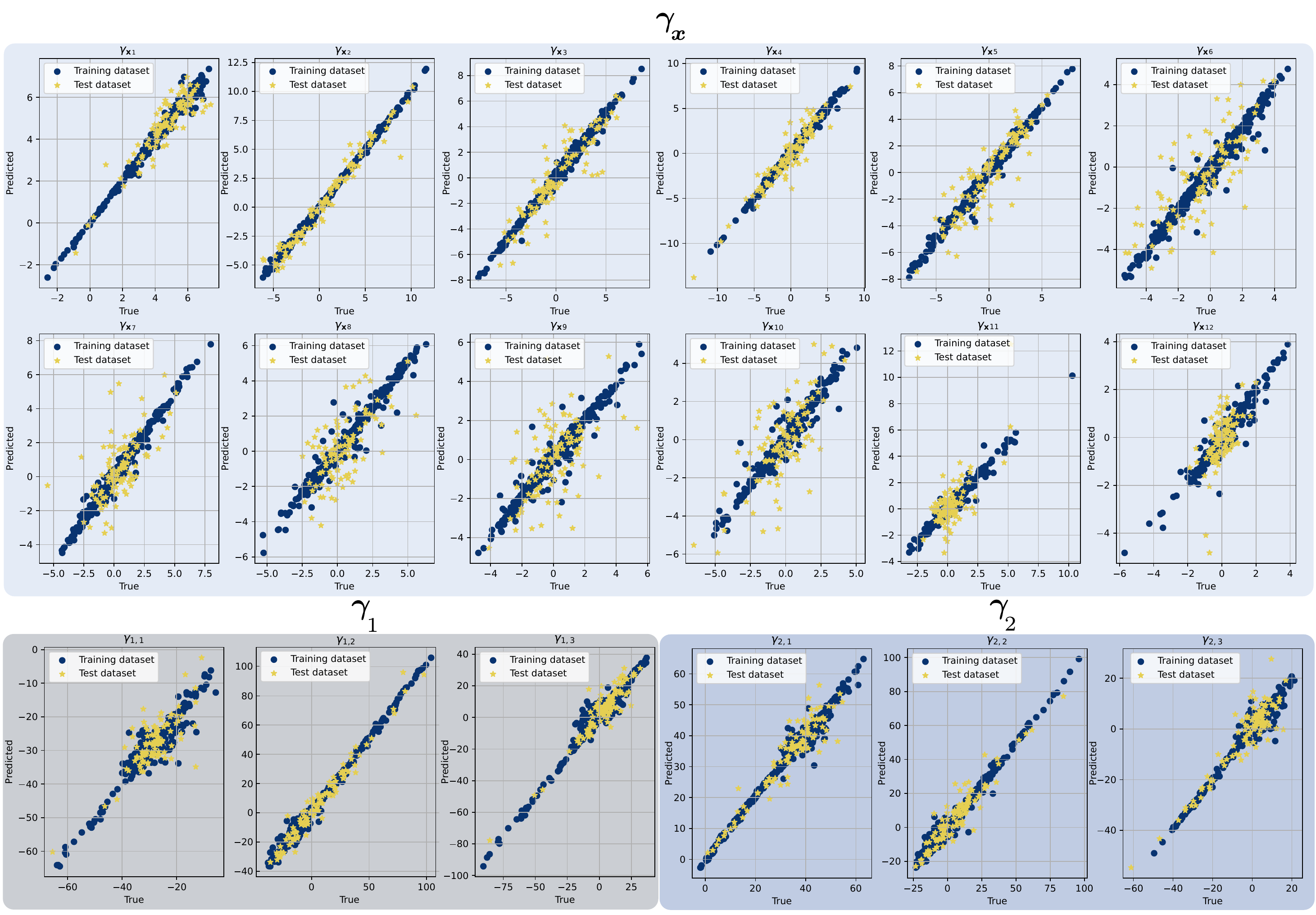}
	\caption{True vs. Predicted latent coefficients ($\boldsymbol{\gamma}_{\boldsymbol{x}}$, $\boldsymbol{\gamma}_{1}$, and $\boldsymbol{\gamma}_{2}$ ) from the three RRAEs models, obtained using the MLP mapping from $\boldsymbol{\alpha}$, for both \textbf{training} and \textbf{test} datasets.}
	\label{fig:MLP_from_Geo_to_GPD_coefficients:Comparison}
\end{figure}

\begin{figure}[bht!]
	\centering              
	\subfloat[\scriptsize{sPGD spatial functions for the first three normalized modes ($F^{i}(\boldsymbol{x})$, $i=1,2,3$).}]{\label{fig:GPD_comparison_x}\includegraphics[width=0.95\textwidth]{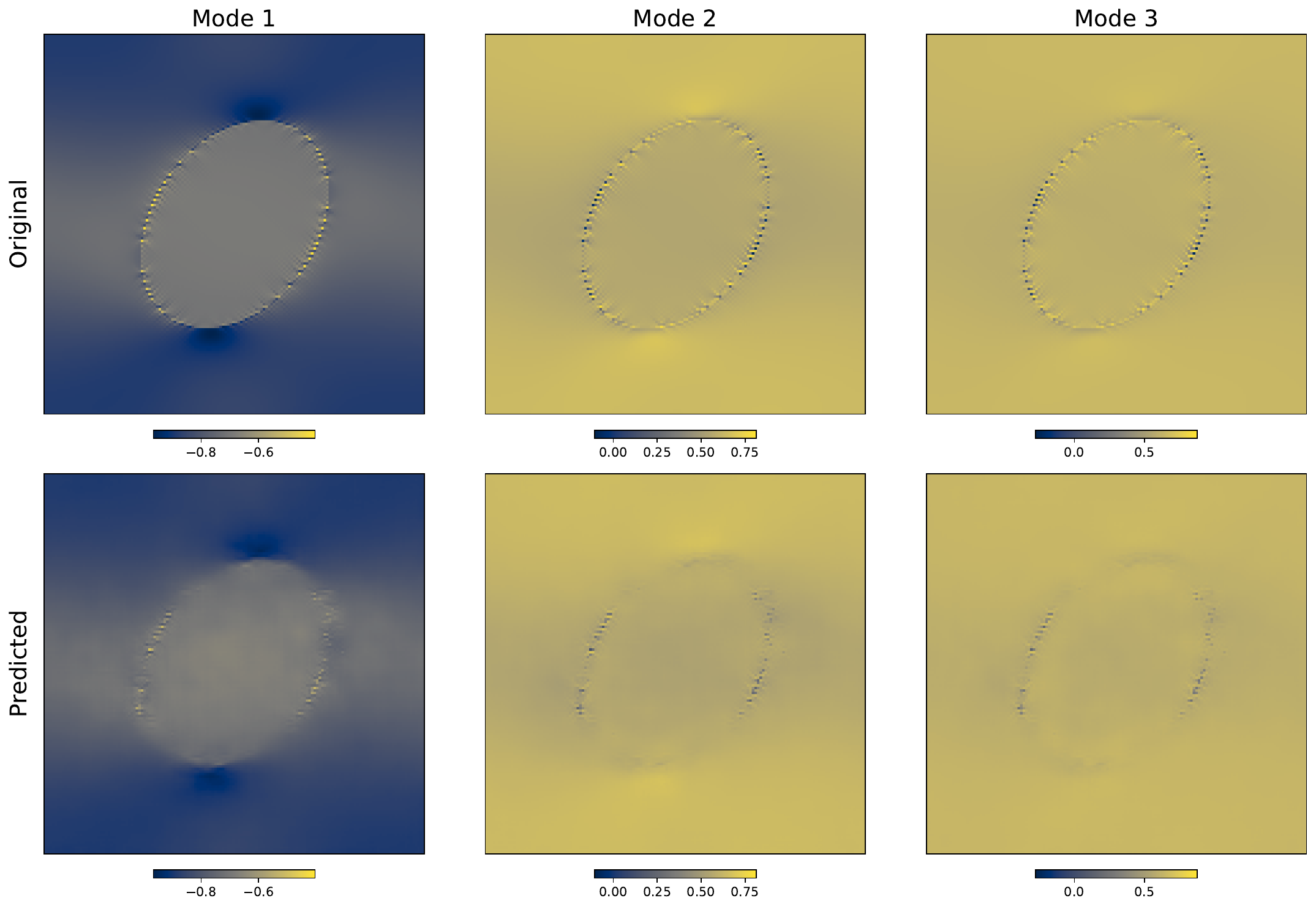}}
	\
	\subfloat[\scriptsize{sPGD one-dimensional functions of $\mu_1$ for the first three modes ($M_{1}^{i}(\mu_{1})$, $i=1,2,3$).}]
	{\label{fig:GPD_comparison_p1}\includegraphics[width=0.45\textwidth]{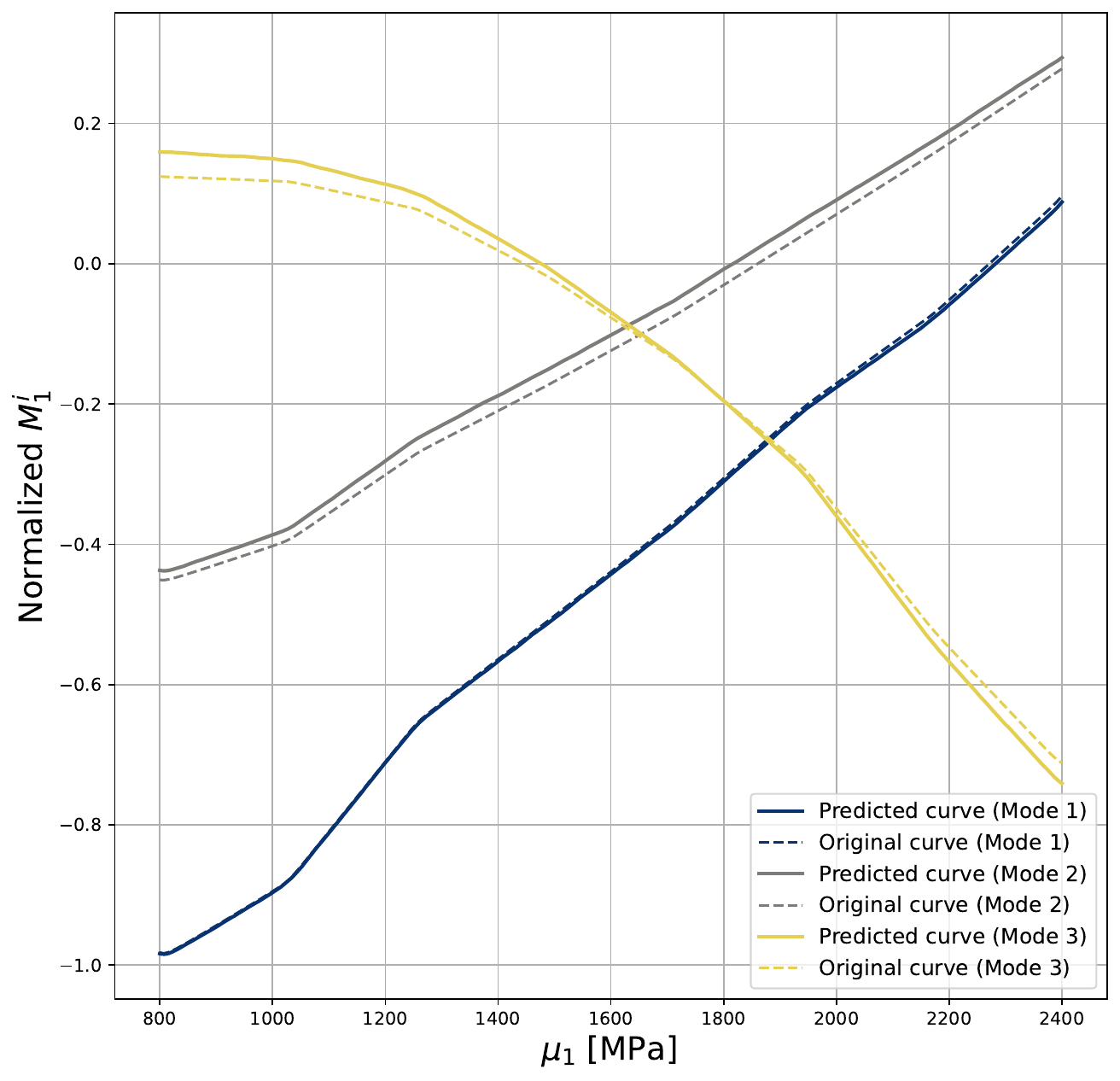}}
	\
	\subfloat[\scriptsize{sPGD one-dimensional functions of $\mu_2$ for the first three modes ($M_{2}^{i}(\mu_{2})$, $i=1,2,3$).}]{\label{fig:GPD_comparison_p2}\includegraphics[width=0.45\textwidth]{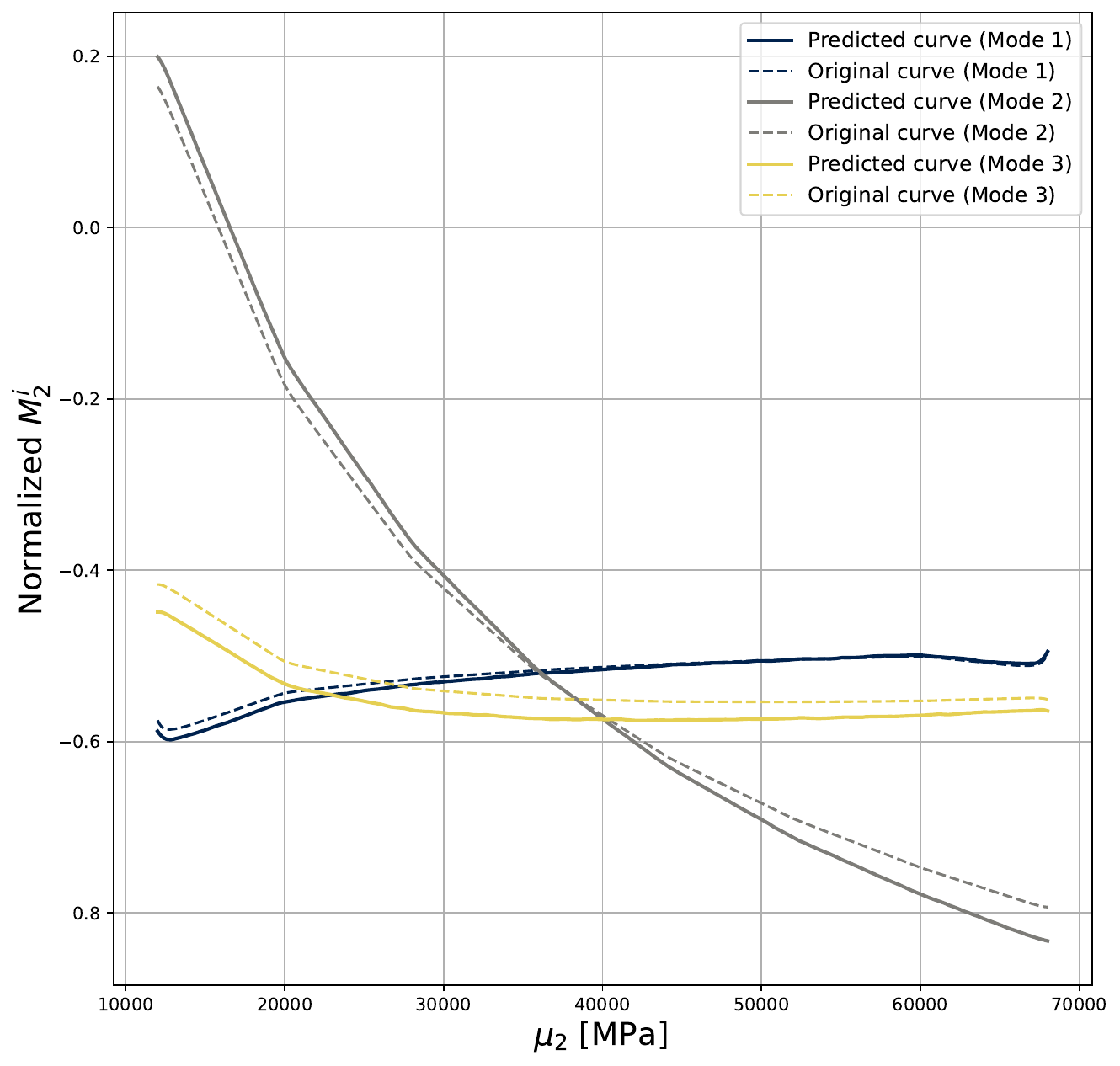}}
	\caption{Comparison of Original and Predicted sPGD functions ($F^{i}(\boldsymbol{x})$, $M_{1}^{i}(\mu_{1})$ and $M_{2}^{i}(\mu_{2})$) for the first three modes obtained from the GPD Framework for a \textbf{test} microstructure.}
	\label{fig:From_Geo_to_GPD_coefficients:Comparison}
\end{figure}

Figure \ref{fig:MLP_from_Geo_to_GPD_coefficients:Comparison} compares the true and predicted latent coefficients $\boldsymbol{\gamma}_{\boldsymbol{x}}$, $\boldsymbol{\gamma}_{1}$, and $\boldsymbol{\gamma}_{2}$ for training points (blue) and test points (yellow). Here, ``true" denotes the coefficients extracted from the trained RRAEs encoders, while ``predicted" denotes the coefficients estimated by the MLP that maps the geometry latent space to the PGD latent spaces. In each case, the coefficients are ordered by decreasing significance due to the truncated SVD, with the leading components capturing the dominant modes of variation. As observed, the leading coefficients are predicted accurately, with both training and test points lying close to the diagonal, indicating a good fit and strong generalization. Higher-order coefficients show greater scatter, particularly for the test set, but still retain a clear correlation with the true values. These findings indicate that the MLPs reliably recover the principal latent structure of the PGD modes. For completeness, Figure \ref{fig:From_Geo_to_GPD_coefficients:Comparison} compares the original and predicted sPGD solutions in separated form, obtained with the GPD framework for a representative test microstructure. The procedure consists of first extracting the geometry latent coefficient $\boldsymbol{\alpha}$ from the trained RRAEs of geometry. This coefficient is then passed through the trained MLPs to predict the latent coefficients associated with each separated function of the PGD approximation. These predicted coefficients are subsequently decoded through the corresponding RRAEs models to reconstruct the functions for all separated variables. The spatial functions (Panel \ref{fig:GPD_comparison_x}) exhibit excellent qualitative agreement between the original and predicted modes, confirming that the main spatial patterns are accurately reproduced. Similarly, the one-dimensional functions with respect to  $\mu_1$ and $\mu_2$ (Panels \ref{fig:GPD_comparison_p1} and \ref{fig:GPD_comparison_p2}) show strong consistency, with predicted curves closely matching the original ones across all three modes. Although minor deviations appear in higher-order modes, the dominant features and overall trends remain well preserved. These results highlight the capability of the GPD framework to reliably approximate both spatial and parametric components of the sPGD solution, ensuring an accurate representation of the most relevant modes in microstructure analysis.

\FloatBarrier
\subsection{Generative design and online construction of multiparametric solutions}
Once the developed GPD framework was trained and validated on the training and test datasets, the next step was to investigate its generative capability. To this end, multiparametric solutions of selected generated geometries were constructed using the GPD framework. Specifically, two microstructures were chosen from the 2000 generated samples, and their latent space coefficients were passed through the trained MLP regression model to predict the corresponding parameters. These predicted parameters were then used within the GPD framework to reconstruct the solutions.

\begin{figure}[hbt!]
	\centering
	\includegraphics[width=0.85\textwidth]{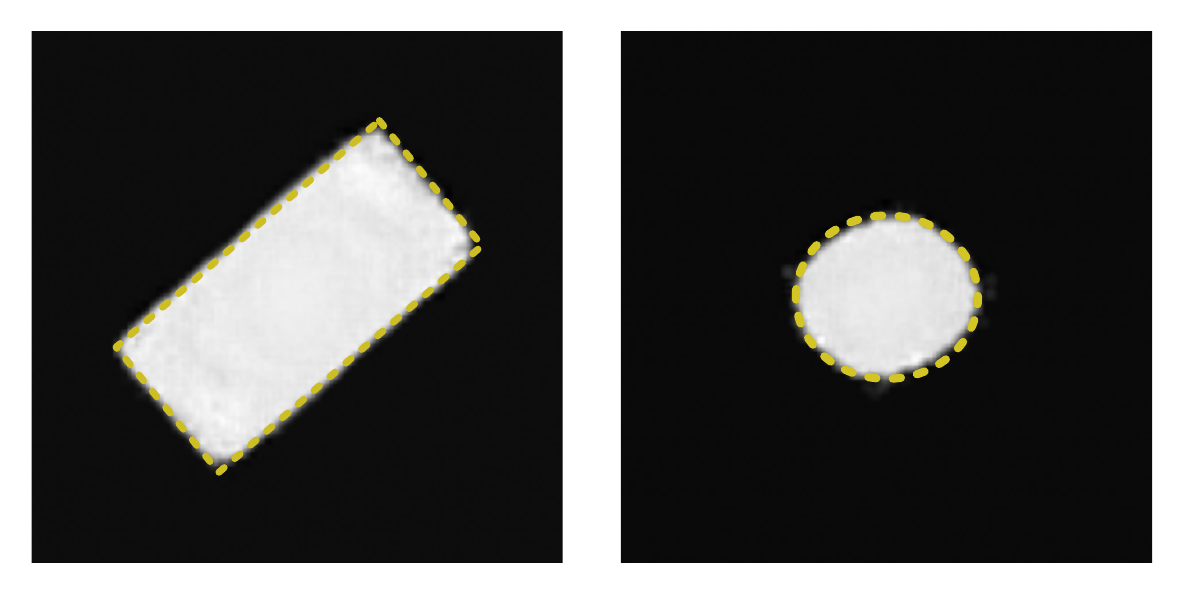}
	\caption{Two representative microstructures out of the 2000 generated using a Gaussian Mixture Model for sampling the latent space $\boldsymbol{\alpha}$, with contours approximating the nominal inclusion geometry.}
	\label{fig:Generated_Microstructure}
\end{figure}

To verify the obtained results, each microstructure morphology was first approximated using the true parameters of the inclusions, allowing the reconstruction of clean geometries without noise (Figure \ref{fig:Generated_Microstructure}). These reconstructed geometries were then meshed and analyzed using FE simulations at each collocation point of the material parameters, as illustrated in Figure \ref{fig:sPGD_grid}. Finally, the sPGD approximation was applied to build the corresponding multiparametric solutions, which were then compared with the predicted results for validation.

\begin{figure}[bht!]
	\centering              
	\subfloat[\scriptsize{sPGD spatial functions for the first three normalized modes ($F^{i}(\boldsymbol{x})$, $i=1,2,3$).}]{\label{fig:Fx_Micro1}\includegraphics[width=0.95\textwidth]{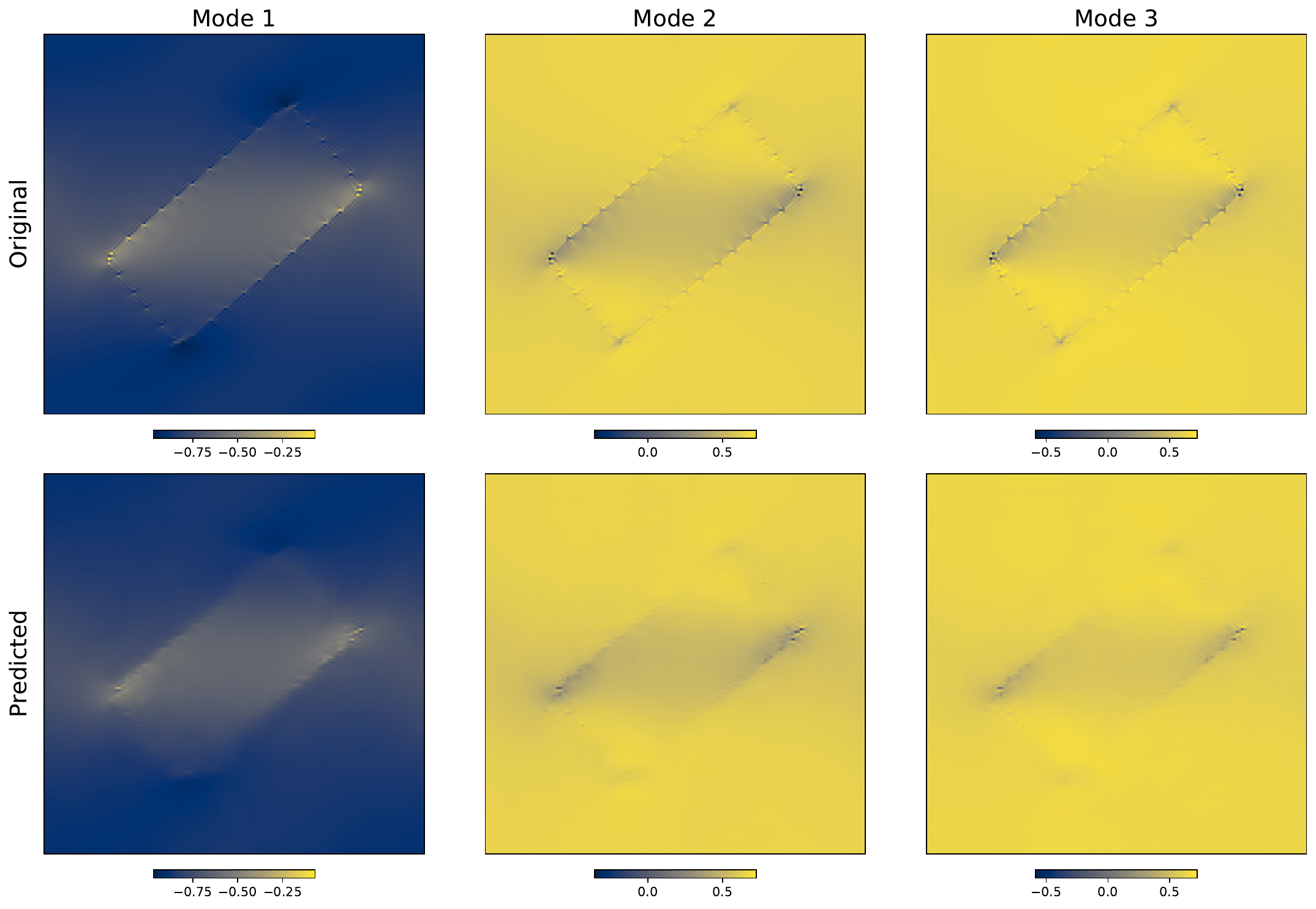}}
	\
	\subfloat[\scriptsize{sPGD one-dimensional functions of $\mu_1$ for the first three modes ($M_{1}^{i}(\mu_{1})$, $i=1,2,3$).}]
	{\label{fig:M1_Micro1}\includegraphics[width=0.45\textwidth]{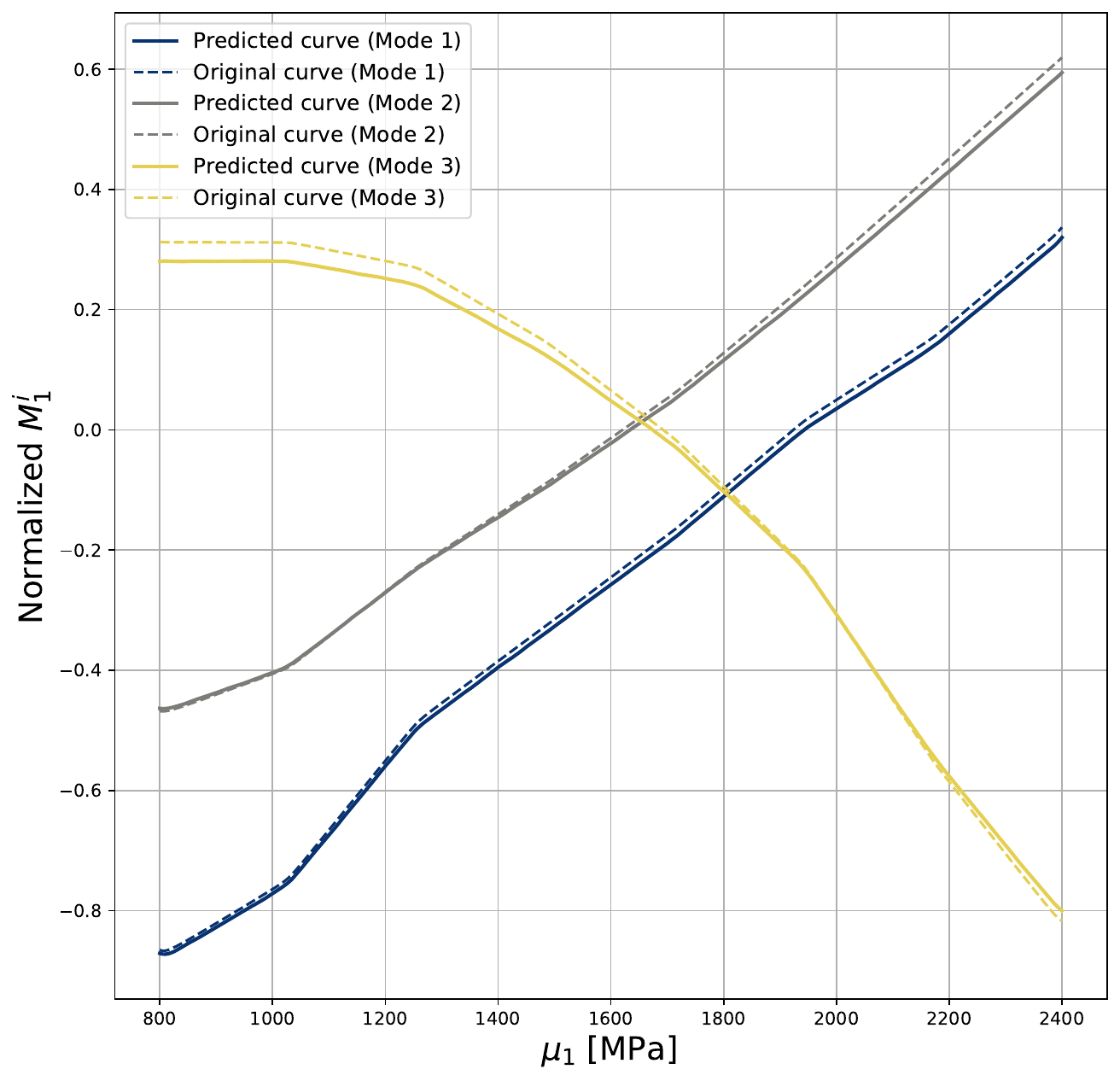}}
	\
	\subfloat[\scriptsize{sPGD one-dimensional functions of $\mu_2$ for the first three modes ($M_{2}^{i}(\mu_{2})$, $i=1,2,3$).}]{\label{fig:M2_Micro1}\includegraphics[width=0.45\textwidth]{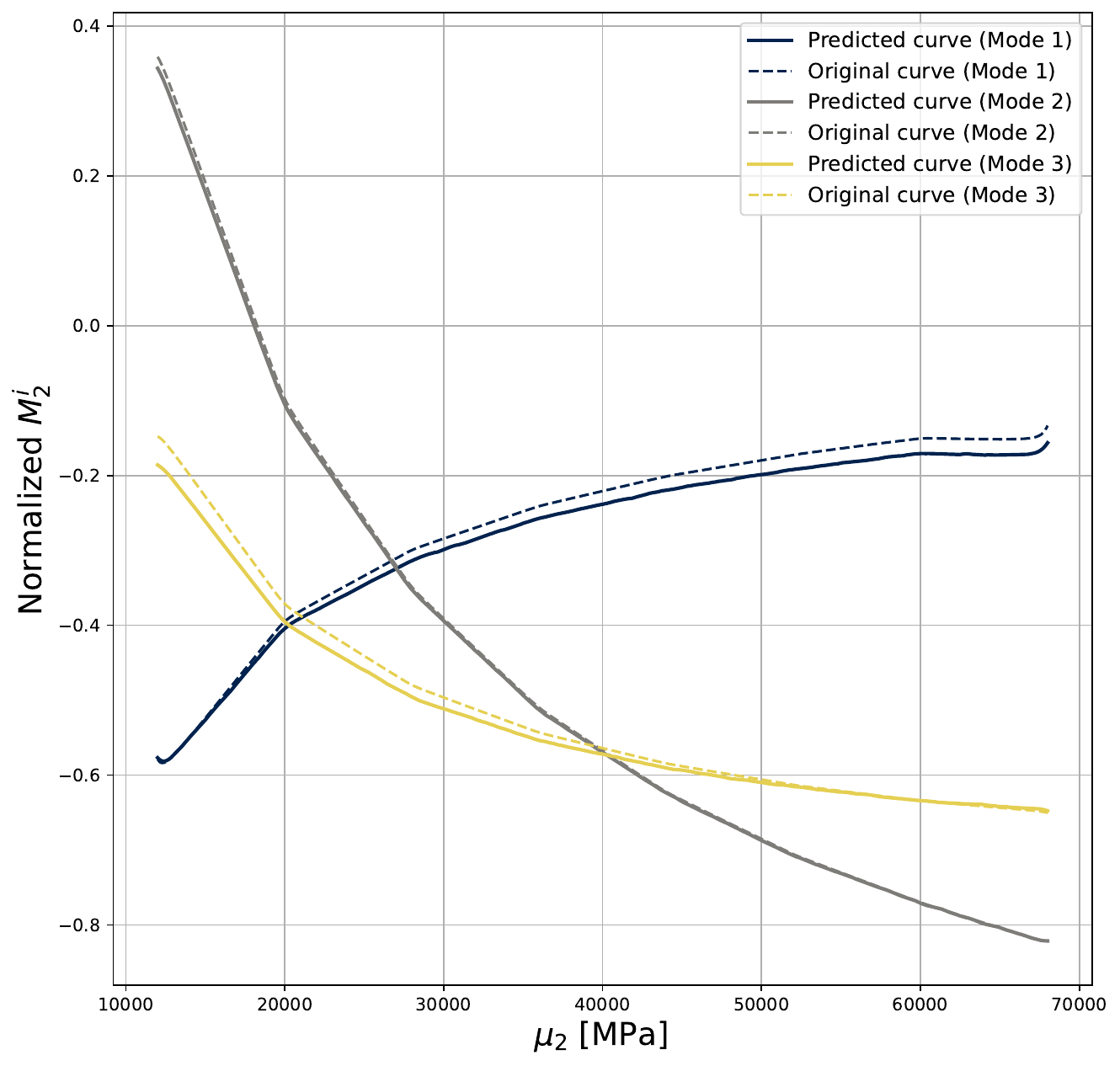}}
	\caption{Comparison of Original and Predicted sPGD functions ($F^{i}(\boldsymbol{x})$, $M_{1}^{i}(\mu_{1})$ and $M_{2}^{i}(\mu_{2})$) for the first three modes obtained from the GPD Framework for a \textbf{newly generated} microstructure (first example)}
	\label{fig:GPD_Micro1}
\end{figure}

\begin{figure}[bht!]
	\centering              
	\subfloat[\scriptsize{sPGD spatial functions for the first three normalized modes ($F^{i}(\boldsymbol{x})$, $i=1,2,3$).}]{\label{fig:Fx_Micro2}\includegraphics[width=0.95\textwidth]{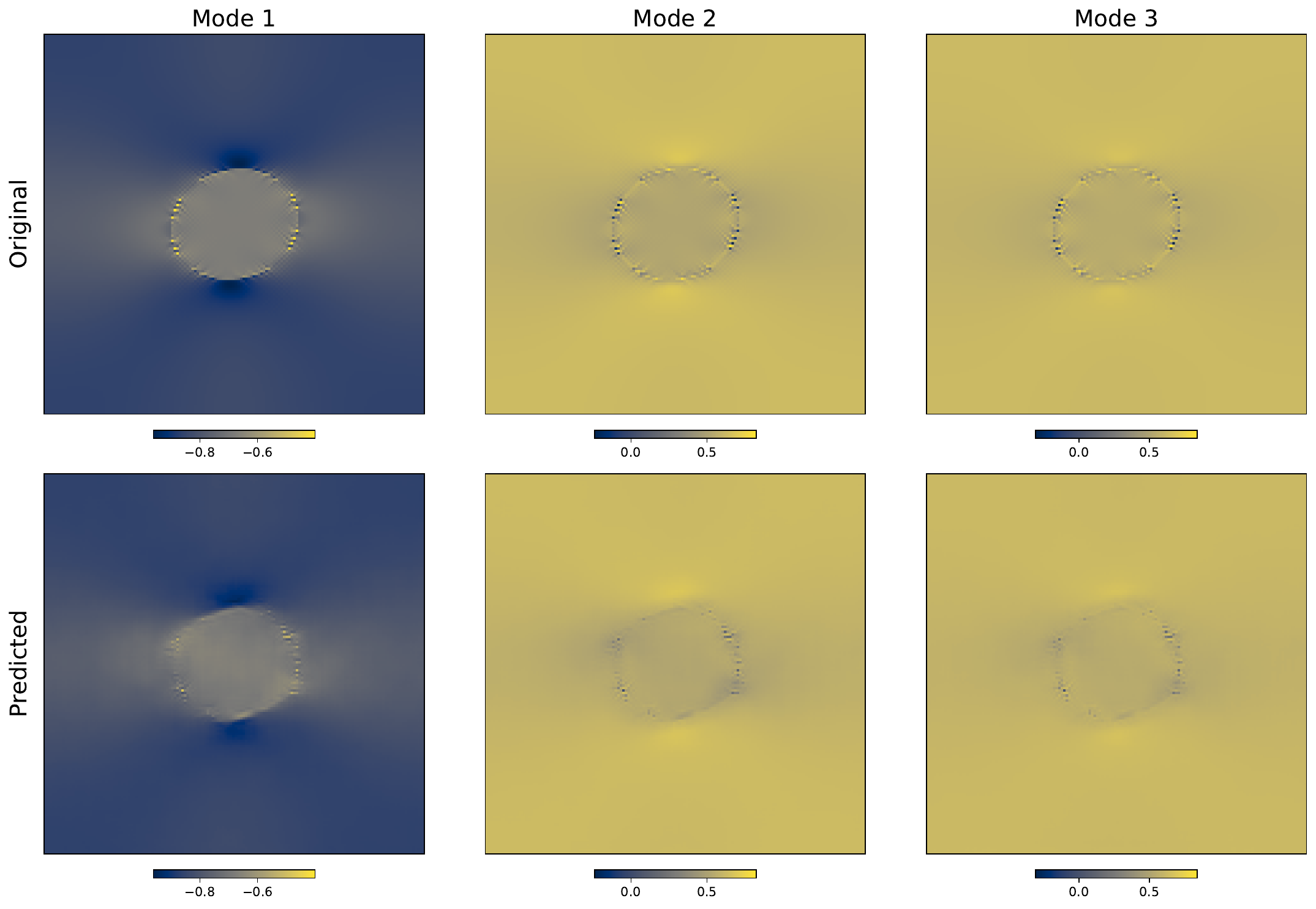}}
	\
	\subfloat[\scriptsize{sPGD one-dimensional functions of $\mu_1$ for the first three modes ($M_{1}^{i}(\mu_{1})$, $i=1,2,3$).}]
	{\label{fig:M1_Micro2}\includegraphics[width=0.45\textwidth]{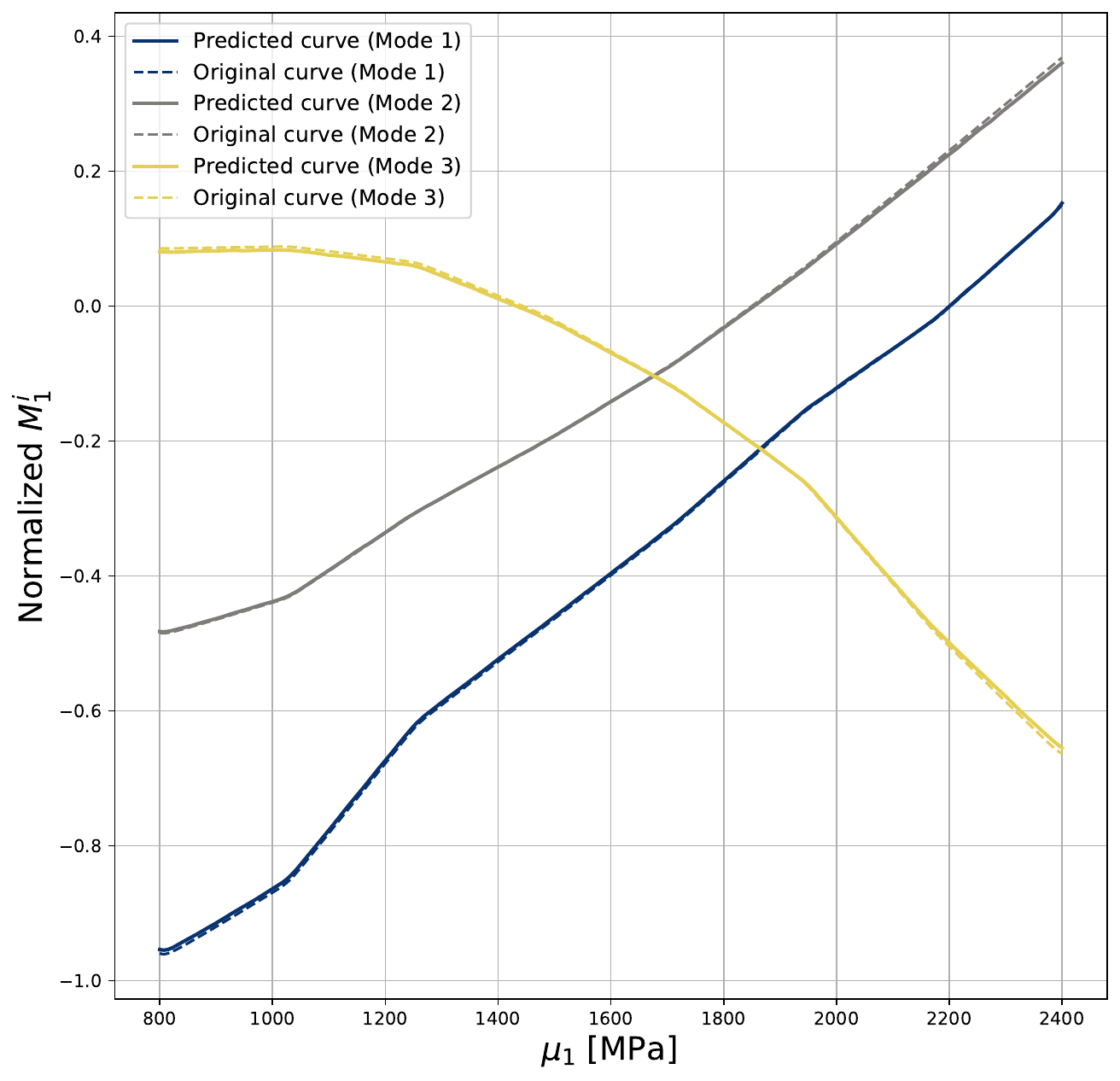}}
	\
	\subfloat[\scriptsize{sPGD one-dimensional functions of $\mu_2$ for the first three modes ($M_{2}^{i}(\mu_{2})$, $i=1,2,3$).}]{\label{fig:M2_Micro2}\includegraphics[width=0.45\textwidth]{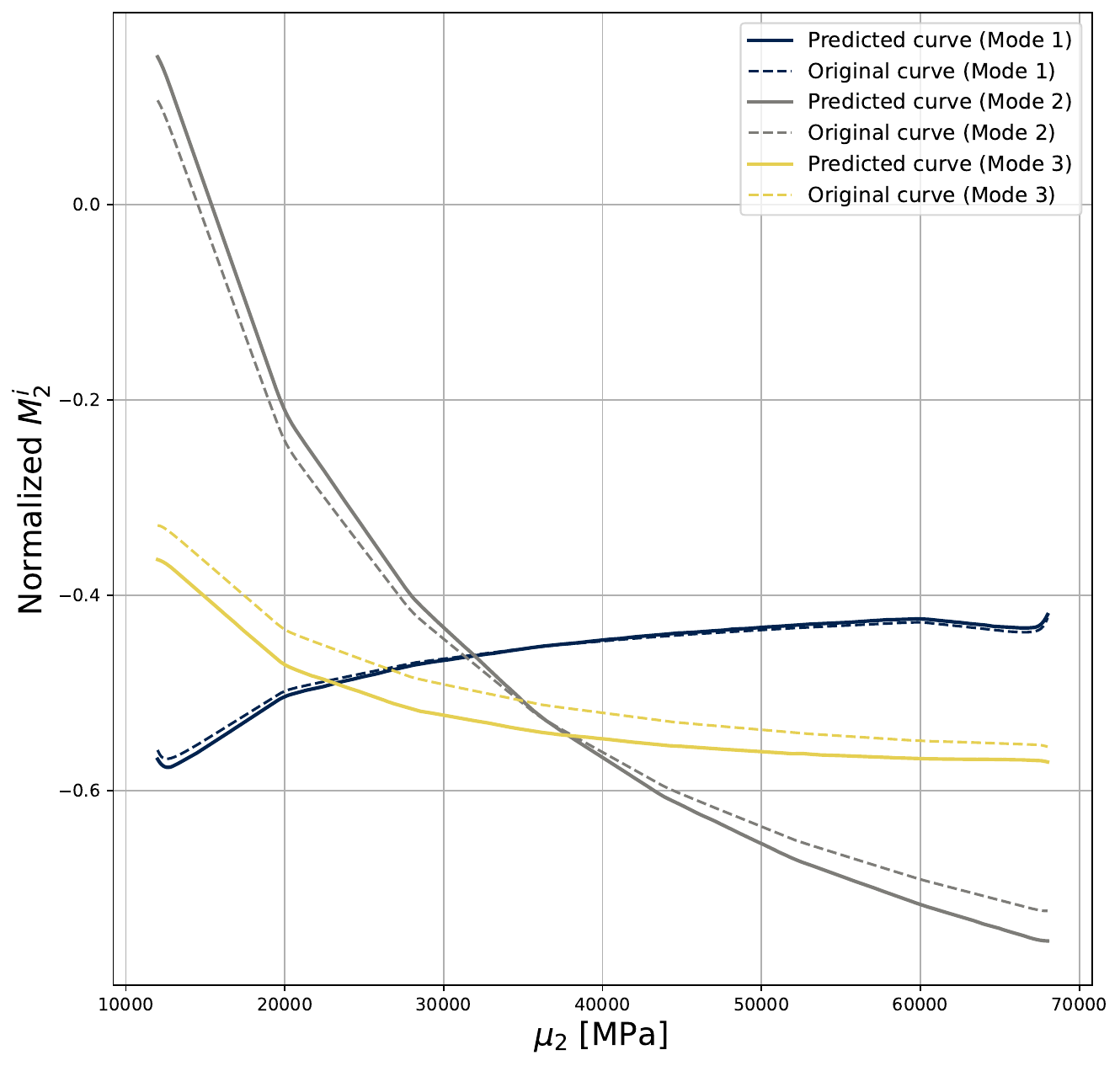}}
	\caption{Comparison of Original and Predicted sPGD functions ($F^{i}(\boldsymbol{x})$, $M_{1}^{i}(\mu_{1})$ and $M_{2}^{i}(\mu_{2})$) for the first three modes obtained from the GPD Framework for a on a \textbf{newly generated} microstructure (second example).}
	\label{fig:GPD_Micro2}
\end{figure}

Figures \ref{fig:GPD_Micro1} and \ref{fig:GPD_Micro2} present a comparison between the predictions obtained through the GPD framework and the direct FE analysis with sPGD approximation for the normalized sPGD modes of two generated microstructures. The results demonstrate the potential of the proposed framework to construct multiparametric solutions for a newly generated microstruture morphology. However, further refinement is needed, particularly at the matrix–inclusion interface where stress concentrations occur. Such improvements could be achieved by enhancing the training of the spatial function loss to better capture interface effects, or by incorporating geometric information as an additional channel during the RRAEs spatial function training.

\FloatBarrier
\section{Conclusion}
\label{sec4: Conclusion}

This work presents the GPD framework and demonstrated its capabilities to adress the dual challenges of generative design and high-dimensional multiparametric solutions. The method integrates Rank-Reduced Autoencoders (RRAEs) with the sparse Proper Generalized Decomposition (sPGD) to tackle complex design problems efficiently. At its core, the framework employs RRAEs to project both design geometries and sPGD solution modes into low-dimensional, and regularized latent spaces, which are then linked through a regression model. This coupling enables a direct prediction of reduced-order solution representations from a given geometry and rapid reconstruction of the full multiparametric solution. Consequently, GPD enables fast exploration and optimization of complex parametric designs, which is especially useful in industrial applications with large parameter spaces and limited high-fidelity data.  Its capacity to support design exploration, optimization, and reduced-order modeling directly contributes to the development of digital and hybrid twins, enabling enhanced predictive modeling and real-time decision-making in engineering workflows.

Qualitative and quantitative evaluations show that the trained RRAE models, for both geometry and PGD modes, effectively compress the original high-dimensional data into a compact set of latent features. These features can be manipulated to establish mappings between different latent spaces or to explore the design space, while still enabling accurate reconstruction of the original data. The results further demonstrate that the GPD framework is capable of successfully generating multiparametric reduced-order solutions, encompassing both spatial and parametric functions, even for previously unseen geometries.

Future work will aim at improving both integration and generalization of the framework. One direction is to train the RRAEs models and the MLP regressors simultaneously in an end-to-end manner, so that the latent spaces remain consistent and better correlated during regression. Another perspective is to develop adaptive versions of the RRAEs that automatically determine the truncation parameter $k_{\max}$, avoiding manual tuning. In addition, the framework will be extended to more complex applications involving three-dimensional geometries and materials with richer behaviors, such as nonlinear or anisotropic responses with multiple parametric inputs, multiphysics couplings, and geometries that are not necessarily confined to a square domain.

\FloatBarrier
\bibliographystyle{elsarticle-harv}
\bibliography{references}

\end{document}